\documentclass[traditabstract]{aa}

\usepackage{threeparttable}
\usepackage{graphicx,rotating,epsfig,color}
\usepackage{times,latexsym,txfonts,natbib}
\usepackage{longtable,lscape,multirow,supertabular,subfigure,caption}

\def\kms{km\,s$^{-1}$}

\def\Mdot{\mbox{\.M}}

\newcommand  \m{\mathrm}

\newcommand  \HII{\,H\,{\footnotesize II} }

\bibpunct{(}{)}{;}{a}{}{,} 
\begin{document}
\title{Blowing in the wind: The dust wave around $\sigma$ Ori AB}
\subtitle{}
\author{B.B. Ochsendorf\inst{\ref{inst1}}, N.L.J. Cox\inst{\ref{inst2}}, S. Krijt \inst{\ref{inst1}}, F. Salgado\inst{\ref{inst1}},  O. Bern\'{e}\inst{\ref{inst4}}, J.P. Bernard\inst{\ref{inst4}}, L. Kaper\inst{\ref{inst3}} \& A.G.G.M. Tielens\inst{\ref{inst1}}}
\institute{Leiden Observatory, Leiden University, P.O. Box 9513, NL-2300 RA, The Netherlands \\ 
\email{ochsendorf@strw.leidenuniv.nl}\label{inst1}
\and
Instituut voor Sterrenkunde, K.U. Leuven, Celestijnenlaan 200D, bus 2401, 3001 Leuven, Belgium\label{inst2}
\and
Sterrenkundig Instituut Anton Pannekoek, University of Amsterdam, Science Park 904, P.O. Box 94249, 1090 GE Amsterdam, The Netherlands\label{inst3}
\and
Universit\'{e} de Toulouse, UPS-OMP, IRAP, 31028 Toulouse, France\label{inst4}}

\abstract{Observations obtained with the Spitzer Space Telescope and the WISE satellite have revealed a prominent arc-like structure at 50" ($\simeq0.1$ pc) from the O9.5V/B0.5V system $\sigma$ Ori AB. We measure a total dust mass of 2.3$\pm$1.5 $\times$ 10$^{-5}$ M$_\odot$. The derived dust-to-gas mass ratio is $\simeq$ 0.29$\pm$0.20.

We attribute this dust structure to the interaction of radiation pressure from the star with dust carried along by the IC 434 photo-evaporative flow of ionized gas from the dark cloud L1630. We have developed a quantitative model for the interaction of a dusty ionized flow with nearby (massive) stars where radiation pressure stalls dust, piling it up at an appreciable distance ($\textgreater$ 0.1 pc), and force it to flow around the star. The model demonstrates that for the conditions in IC 434, the gas will decouple from the dust and will keep its original flow lines. Hence, we argue that this dust structure is the first example of a dust wave created by a massive star moving through the interstellar medium. Our model shows that for higher gas densities, coupling is more efficient and a bow wave will form, containing both dust and gas.

Our model describes the physics of dust waves and bow waves and quantitatively reproduces the optical depth profile at 70 $\mu$m. Dust waves (and bow waves) stratify dust grains according to their radiation pressure opacity, which reflects the size distribution and composition of the grain material. It is found that in the particular case of $\sigma$ Ori AB, dust is able to survive inside the ionized region.  Comparison of our model results with observations implies that dust-gas coupling through Coulomb interaction is less important than previously thought, challenging our understanding of grain dynamics in hot, ionized regions of space. 

We describe the difference between dust (and bow) waves and classical bow shocks created by the interaction of a stellar wind with the interstellar medium. The results show that for late O-type stars with weak stellar winds, the stand-off distance of the resulting bow shock is very close to the star, well within the location of the dust wave. In general, we conclude that dust waves and bow waves should be common around stars showing the weak-wind phenomenon, i.e., stars with log$(L/L_{\odot})$ $\textless$ 5.2, and that these structures are best observed at mid-IR to FIR wavelengths, depending on the stellar spectral type. In particular, dust waves and bow waves are most efficiently formed around weak-wind stars moving through a high density medium. Moreover, they provide a unique opportunity to study the direct interaction between a (massive) star and its immediate surroundings.}

\authorrunning{B.B. Ochsendorf et al.}
\titlerunning{The dust wave around $\sigma$ Ori AB}
\maketitle

\section{Introduction}\label{sec:intro}

The initial expansion of \HII regions is driven by rapid ionization around a young, massive star. Once the \HII region is established, the overpressure of the ionized region will drive the expansion as the ionizing flux of the star dilutes with distance, sweeping up neutral gas in a shell. Dust inside the \HII region will absorb ionizing flux and re-emit in the IR, stalling the expansion of the ionized material  \citep{petrosian_1973,draine_2011}. When the ionization front breaks out of the molecular cloud, a phase of rapid expansion of the ionized gas into the surrounding low density medium commences: the so-called champagne flow phase \citep{tenorio_tagle_1979, bedijn_1981, yorke_1986}.

Massive stars have strong winds ($\Mdot$ $\sim$ 10$^{-8}$ - 10$^{-6}$ M$_\odot$ yr$^{-1}$, $v_\infty$ $\simeq$ 1000 - 2500 km s$^{-1}$; \citet{kudritzki_2000}). If these stars move supersonically with respect to their environment, a bow shock will be created at the stand-off distance from the moving star where the ram pressure and momentum flux of the wind and the interstellar medium balance \citep{van_buren_1990, maclow_1991}. The swept-up material in the bow shock is heated by the radiation of the OB star, making these structures visible in either shocked gas \citep{brown_2005, kaper_1997} or warm dust radiating at infrared wavelengths \citep{van_buren_1988}. The Infrared Astronomical Satellite (IRAS; \citet{neugebauer_1984}) all-sky survey detected many extended arc-like structures associated with OB-runaway stars, revealing that bow shocks are ubiquitous around stars with powerful stellar winds \citep{van_buren_1995,peri_2012}. However, recent observations indicate that many late O-type dwarfs have mass-loss rates which are significantly lower than predicted by theory \citep{bouret_2003,martins_2004}, which raises questions whether the observed scale sizes of bow shocks around stars that display the weak-wind phenomenon can be accounted for by a wind-driven bow shock model. It has already been proposed by \citet{van_buren_1988} that radiation pressure driven bow waves are expected around stars with a low wind momentum flux. Until now, no such structure has been detected in the interstellar medium (ISM).

The IC 434 emission nebula is probably an evolved \HII region, where the ionizing population of the large $\sigma$ Orionis cluster has cleared away its immediate surroundings. Currently, the ionization front is eating its way into the L1630 molecular cloud and the ionized material is streaming into the \HII region. The central component is a five-star system, found approximately one degree below Altinak, the easternmost star in Orion's Belt. This region also contains the characteristic Horsehead Nebula (Barnard 33), emerging as a dark nebula out of the large L1630 molecular cloud. \citet{caballero_2008} has measured a distances of 334$^{+25}_{-22}$ pc, although reported distances vary up to $\sim$500 pc, as determined from colour-magnitude diagrams \citep{caballero_2007b}. Previous studies have revealed a vast population of lower-mass stars and brown dwarfs belonging to the larger $\sigma$ Ori open cluster \citep{bejar_2011,caballero_2007a}. The dominant ionizing component of the system, $\sigma$\,Ori AB,  is a 3 Myr old \citep{caballero_2008} close binary of spectral type O9.5V and B0.5V with a third massive companion \citep{simon_diaz_2011}. The ionizing flux from the central stars illuminates IC 434 together with the mane of the Horsehead nebula. The presence of a 24 $\mu$m infrared arc-like structure near $\sigma$ Ori has been noted previously by \citet{caballero_2008b} (their Fig. 1, as well as Fig. 2 from \citet{hernandez_2007}), who designated the brightest knot as \object{$\sigma$ Ori IRS2}. To the best of our knowledge, we present here the first detailed study of this unique and conspicuous infrared source.

We argue that the arc structure engulfing $\sigma$\,Ori~AB presents the first detection of a {\em dust wave} in the ISM, where radiation pressure has stalled the dust which was carried along by a photo-evaporation flow of ionized gas. In Sec. \ref{sec:observations}, we present the observations and how we processed the data; Sec. \ref{sec:basic} reviews the stellar properties and we derive local physical parameters; in Sec. \ref{sec:structure} the observables and the global structure are presented; in Sec. \ref{sec:flow} we propose the dust wave scenario to explain the observations; results are shown in Sec. \ref{sec:results} which we will discuss in Sec. \ref{sec:discussion}. We conclude in Sec. \ref{sec:conclusions}.

\section{Observations}\label{sec:observations}

\paragraph{{\em IR photometry}} \hspace{0pt} \\
Infrared (IR) photometry of the $\sigma$ Ori cluster combines data from the {\it Wide-field Infrared Survey Explorer} (WISE; \citet{wright_2010}), the Infrared Array Camera (IRAC; \citet{fazio_2004}) and the Multiband Imaging Photometer (MIPS; \citet{rieke_2004}) on board of the {\it Spitzer Space Telescope} and the Photodetector Array Camera and Spectrometer (PACS; \citet{poglitsch_2010}) on the {\it Herschel Space Telescope}. The WISE atlas data were extracted from the All-Sky Data release and were mosaiced using Montage. IRAC data were observed in March 2004 as part of program ID 37 and were taken from the Spitzer Science Center archive as a post-Basic Calibrated Data (post-BCD) product. The post-BCD image with a 1$^{\degr}$$\times$0.8$^{\degr}$ field-of-view (FOV) combines BCD images of 30 seconds integration time each and was found to be of good quality. The MIPS 24 $\mu$m post-BCD data was also of sufficient quality and give a 0.75$^{\degr}$$\times$1.5$^{\degr}$ FOV in medium scan mapping mode with 160$\arcsec$ steps. PACS 70 $\mu$m photometry was taken from the Herschel Science Archive (HSA) from the Gould Belt survey \citep{andre_2010} with observation IDs 1342215984 and 1342215985. These PACS data were obtained in PACS/SPIRE parallel mode at high scanning speed (60$\arcsec$ s$^{-1}$). The nominal full width at half maximum (FWHM) of the point spread function (PSF) for this type of observations is 5.9$\arcsec$ $\times$ 12.2$\arcsec$ for the blue (70 $\mu$m) channel. HIPE v.8 \citep{ott_2010} and Scanamorphos v.15 \citep{roussel_2010} were used to make maps of the IC 434 region after which the data was inter-calibrated with IRAS 70 $\mu$m data. Contamination by zodiacal light was subtracted using the SPOT background estimator, based on the COBE/DIRBE model \citep{kelsall_1998}.

\paragraph{{\em H$\alpha$ imaging}} \hspace{0pt} \\
H$\alpha$ narrow band data from the {\em Hubble Space Telescope} (HST), the Mosaic 1 wide field imager on Kitt Peak National Observatoy (KPNO) and the Southern H-Alpha Sky Survey Atlas (SHASSA) are used in this study. The HST image of the Horsehead was taken as part of the Hubble Heritage program (PI: K. Noll) and is used as a calibrator for the KPNO image. The calibrated KPNO image offers H$\alpha$ data at a high angular resolution (0.26 " pix$^{-1}$), but does not extend towards $\sigma$ Ori~AB. Because of this, we complement our H$\alpha$ observations with data from the SHASSA mission. SHASSA provides a complete coverage of the northern hemisphere at a 0.8 arcminute resolution.

\paragraph{{\em Spitzer/IRS spectral mapping}} \hspace{0pt} \\
Spitzer/IRS \citep{houck_2004} spectral maps from the Spitzer Science Center Archive were obtained as part of the SPEC\HII program (PI: C. Joblin) in April 2008. Both Long-Low modules (LL) were used from the observations, which cover the 16-35 $\mu$m wavelength region at an angular resolution of 5$\arcsec$/pixel. In addition, a dedicated offset position was taken at an IRAS dark position. Spectral cubes were built using CUBISM \citep{smith_2007}. CUBISM uses the 2D BCD data and allows for standard reduction operations such as sky subtraction and bad pixel clipping. In addition, slit and aperture loss correction functions appropriate for extended source calibration are applied and statistical errors originating from the IRS pipeline are propagated through CUBISM.

\begin{figure*}
\centering
\includegraphics[width=17cm]{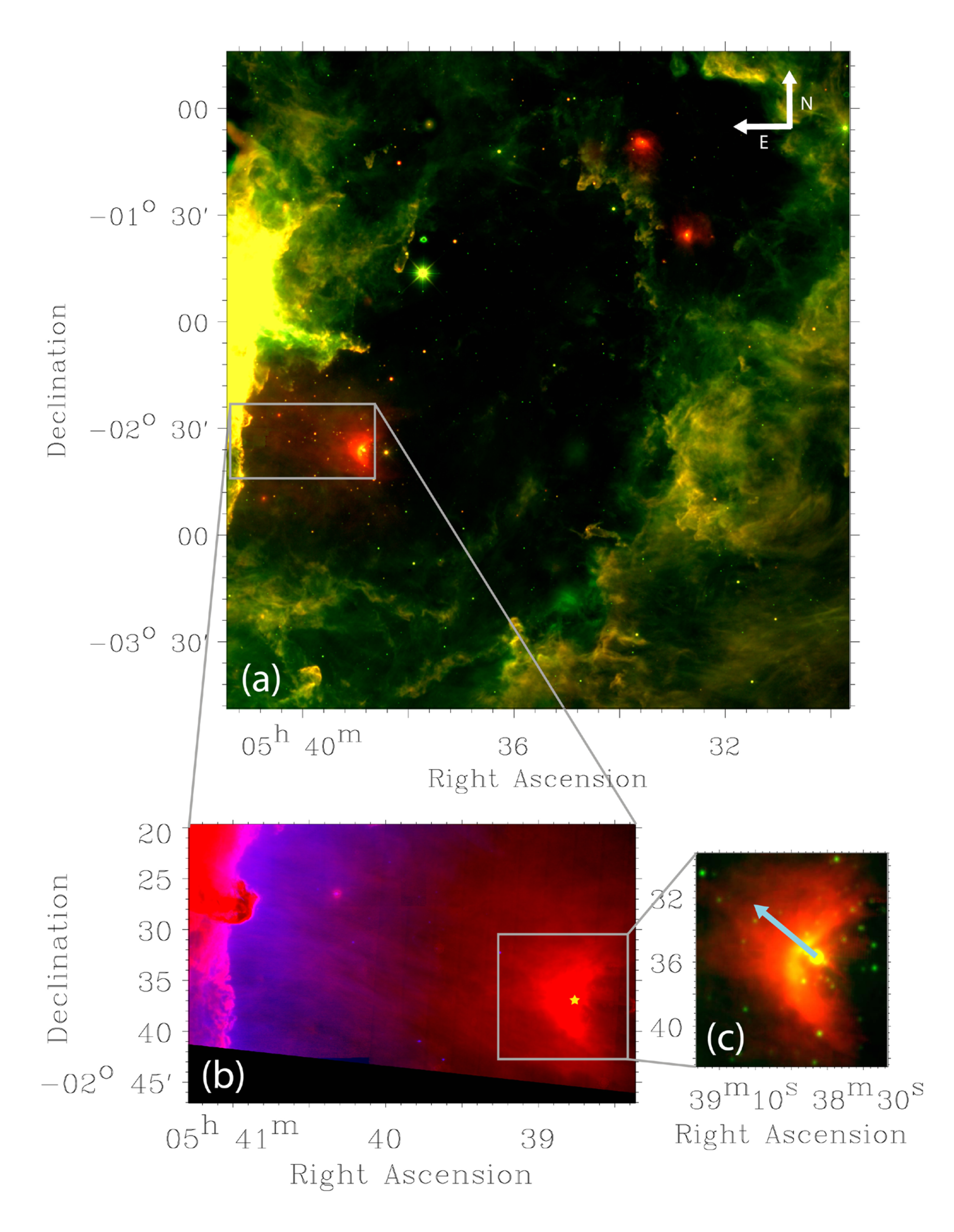} 
\caption{{\bf (a)}: Mid-IR view of the IC 434 region. The total field-of-view is 3.1$^{\degr}$ $\times$ 2.9$^{\degr}$. Green is the WISE-3 12 $\mu$m band; red is the WISE-4 22 $\mu$m band. {\bf (b)}:  Blown-up part of the upper image with different scaling to accentuate the extended emission around $\sigma$ Ori AB. The Horsehead nebula is located at the top left. Here, blue represents H$\alpha$ taken with the KPNO 4m telescope, whereas red is MIPS 24 $\mu$m emission. The KPNO image does not extend over the entire field of view but cuts off 6$\arcmin$ eastwards of $\sigma$ Ori~AB. {\bf (c)}: Close up of the environs of $\sigma$ Ori AB. Green is WISE-3, while red is MIPS 24 $\mu$m. Overplotted is the proper motion vector which displays the movement in the plane of the sky.}
\label{fig:mosaic_withzoom}
\end{figure*}

\section{Stellar properties and local physical conditions}\label{sec:basic}

Figure \ref{fig:mosaic_withzoom} displays the mid-IR view of the IC 434 complex. Although diffuse structures are seen inside the \HII region, the image shows a  large cleared-out region with $\sigma$ Ori AB slightly displaced from its center. The image shows that $\sigma$ Ori AB is indeed the driving ionizing source of the region as numerous pillars of material, which are currently being photo-evaporated, are directed towards the star. The lower panels in Fig. \ref{fig:mosaic_withzoom} zooms into our region of interest: the $\sigma$ Ori AB - L1630 region. Radiation from the star is ionizing the boundary of the molecular cloud, launching the material into the \HII region. Extended emission at 24 $\mu$m and an arc structure, more pronounced at 12 $\mu$m, reveals the interaction of the ionized flow with either a stellar wind or radiation pressure emanating from $\sigma$ Ori AB. 

\subsection{Basic stellar properties and space velocities}\label{sec:spacevel}

Table~\ref{tab:stellar} lists basic stellar properties for $\sigma$\,Ori~AB. We calculate the space velocity of $\sigma$\,Ori~AB with respect to the Local Standard of Rest (LSR). Using these parameters, $\sigma$\,Ori AB has a space velocity of 15.1~\kms, with a position angle of 49.9\degr. For comparison, $\sigma$\,Ori D and $\sigma$\,Ori E have space velocities of 18.5 and 13.1 km s$^{-1}$, respectively. 

The direction of the space velocity of $\sigma$ Ori AB (see Sec. \ref{sec:spacevel}) suggests that it is moving towards the molecular cloud, increasing the relative velocity between the flow and the star. We note that it is possible that the observed space velocity of the $\sigma$ Ori AB (as a member of the $\sigma$ Ori cluster) represents the velocity of the entire region with respect to the Local Standard of Rest (LSR), which would imply that $\sigma$ Ori AB is at rest with the L1630 molecular cloud. A comparison of radial velocities (with respect to the LSR) $v_\m{LSR}$ of members of the $\sigma$ Ori cluster (i.e., $\sigma$ Ori AB,  $\sigma$ Ori D and Ori $\sigma$ E) and the L1630 cloud reveal similar velocities: we calculate a mean $v_\m{LSR}$ of 13 km s$^{-1}$ for the $\sigma$ Ori cluster members, whereas the L1630 cloud moves at a velocity of $v_\m{LSR}$ $\sim$ 10 km s$^{-1}$ as seen through molecular observations \citep{maddalena_1986,gibb_1995}. Surely, a comparison of radial velocities alone will not resolve the question whether both structures are moving together or not, as these measurements only represent one component of the true space velocity. In this work, we will use the measured space velocity of $\sigma$ Ori AB as the relative velocity between the cloud and the star, but we will address the implications on our conclusions if the two structures were at rest with one another.

\begin{table}
\begin{threeparttable}[b]
\caption{Stellar parameters for Sigma Ori AB (also known as 48 Ori, HD 37468, and HIP 26549) and calculated space velocity parameters. $L$ is the luminosity of the star; $Q_\mathrm{0}$ is the ionizing photon flux; $v_\mathrm{rad}$ is the radial velocity; $\mu_\alpha$ and $\mu_\delta$ are the proper motions in RA and DEC, respectively; $d$ is the measured distance; $\Mdot$ is the mass-loss rate. The proper motion position angle $\theta$ is measured from north to east; $i$ is the inclination angle with respect to the sky and $v_\star$ is the total space velocity of $\sigma$\,Ori~AB. }
\label{tab:stellar}
\begin{tabular}{lll}\hline\hline
RA (J2000)	& 05:38:44.768 & \citet{caballero_2007a} \\
DEC (J2000)	& -02:36:00.25 		&     \\
Spectral type   & O9.5V	+ B0.5 V			& \citet{hoffleit_1982}  \\
$L$ (log($L_\star$/$L_\odot$)) & 4.78 & \citet{lee_1968} \\
log $Q_\mathrm{0}$ (s$^{-1}$) & 47.88 & \citet{martins_2005} \\ 
$\mu_\alpha$ (mas yr$^{-1}$)	& 4.61 $\pm$ 0.88 & \citet{perryman_1997}\footnotemark \\
$\mu_\delta$ (mas yr$^{-1}$)	& -0.4 $\pm$ 0.53		& \citet{perryman_1997} \\
$\mu_\alpha$ (km s$^{-1}$)	& 6.8 &  \\
$\mu_\delta$ (km s$^{-1}$)	& 5.7	 &  \\
$v_\mathrm{rad}$(LSR) (km s$^{-1}$) &  11.45	& \citet{caballero_2008} \\
$d$ (pc)	& 334$^{+25}_{-22}$			& \citet{caballero_2008} \\
$\Mdot\,$ (M$_\odot$ yr$^{-1}$)	& 8.0 $\times$ 10$^{-8}$ & \citet{howarth_1989} \\
	& 2.0 $\times$ 10$^{-10}$ & \citet{najarro_2011} \\
$v_\infty$ (km~s$^{-1}$)		& 1060 		& \citet{howarth_1989} \\
& 1500		& \citet{najarro_2011} \\ \hline
$\theta$ ($\degr$)			& 49.9 & This work	\\
$i$ ($\degr$) & 49.4 &	\\
$v_\star$ (km s$^{-1}$)			& 15.1	& \\\hline
\end{tabular}
\begin{tablenotes}
    \item[1]Values for proper motion published in \citet{van_leeuwen_2007} show a large discrepancy between different members $\sigma$ Ori central cluster. Therefore, we adopt proper motion parameters from \citet{perryman_1997}. In this case, proper motions between $\sigma$ Ori AB are similar to the values for $\sigma$\,Ori D and $\sigma$\,Ori E listed in \citet{caballero_2007a}.
  \end{tablenotes}
 \end{threeparttable}
\end{table}

\subsection{Wind properties}
The winds from early O-type stars and early B-type supergiants are well described by the mass-loss recipe from \citet{vink_2000}. However, the accuracy of this recipe has been questioned for stars with luminosity $\mathrm{log}(L/L_\odot)$ $\textless$ 5.2, in particular because of wind clumping and the weak-wind problem as described in \citet{najarro_2011,puls_2008}, but see \citet{huenemoerder_2012}. This is because traditional mass-loss indicators (UV, H$\alpha$) become insensitive at low mass-loss rates. Thus far, there are wind parameters for only a handful of late O-dwarf stars \citep{bouret_2003,martins_2004,najarro_2011}. Table~\ref{tab:stellar} includes two different values for the mass-loss rate $\Mdot$ and the terminal wind velocity $v_\infty$ of $\sigma$ Ori AB. Earlier measurements by \citet{howarth_1989} were interpreted in terms of a much higher momentum flux. However, it is now well appreciated that the diagnostic use of UV wind lines is limited \citep{martins_2005b,puls_2008}. \citet{najarro_2011} categorized $\sigma$\,Ori AB as a weak-wind candidate after careful modeling of spectral lines in the L-band. We believe that the latter value is closer to the true value given the extreme sensitivity of the Br$\alpha$ line flux for very low mass-loss rates \citep{auer_1969,najarro_1998,lenorzer_2004}. The model presented in this work naturally follows from $\sigma$ Ori AB being a weak-wind object; however, we will discuss the implications when one would use a higher momentum flux in Sec. \ref{sec:bowshock} and in Appendix \ref{sec:appendix}.

\subsection{Photo-evaporation flow}\label{sec:emissionmeasure}

The IC434 \HII region is a clear example of a flow of ionized gas, which is initiated when the ionization front breaks out of a dense confining molecular cloud environment into a surrounding tenuous medium. The density contrast will drive a strong isothermal shock into the low density region, while a rarefaction wave will plough into the ionized cloud gas. Material ionized at the edge of the molecular cloud is rapidly accelerated and driven into the low density region, starting the onset of a flow of ionized gas (a {\em champagne flow}; \citet{tenorio_tagle_1979, tielens_2005}). In the case of IC 434, it is the density contrast between the tenuous \HII region and the high density L1630 cloud that has set up the flow of ionized gas towards $\sigma$ Ori AB.

\begin{figure}
\includegraphics[width=9cm]{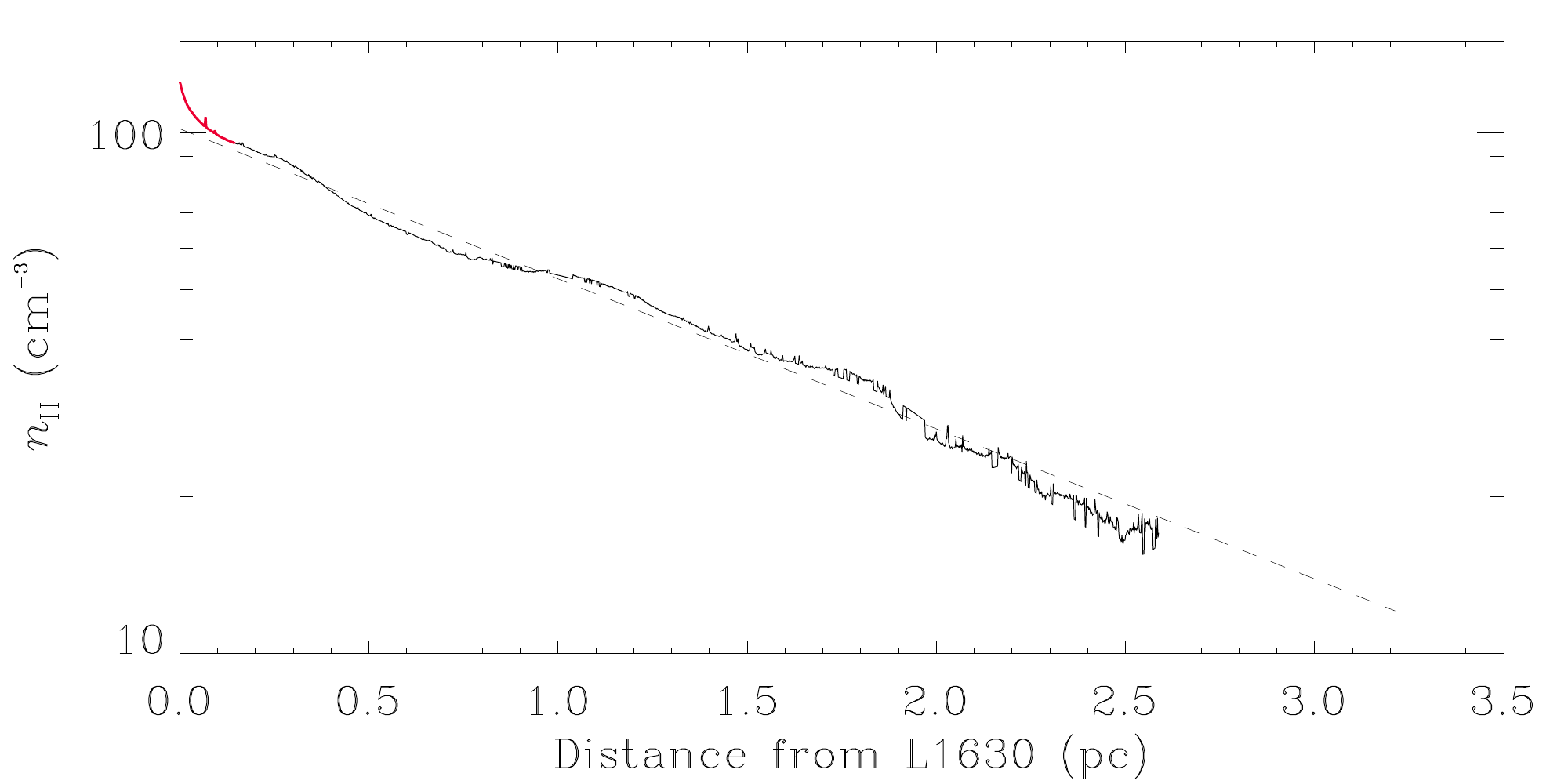} 
\caption{Density profile of the ionized photo-evaporation flow inside IC 434. The contribution of the Horsehead is plotted in red, which is only a small disturbance at the start of the flow. Small artifacts in the observed density profile persist after subtraction of background/foreground stellar features not associated with the flow. The observed profile reveals an exponential density gradient and is fitted using an appropriate density law (dashed line; see text). $\sigma$ Ori~AB is located at a distance of 3.2 pc.}
\label{fig:halphacut}
\end{figure}

Figure \ref{fig:halphacut} shows the root mean square (RMS) electron density derived from the observed H$\alpha$ emission measure (EM), assuming a path length of the emitting region along the line of sight. The EM is defined as $\int n_\mathrm{e}^2\mathrm{d}{l}$, where $n_\mathrm{e}$ is the electron density and $l$ the path length of the emitting region. We consider the \HII region to be fully ionized, i.e., $n_\mathrm{e}$ = $n_\mathrm{H}$, with $n_\mathrm{H}$ the hydrogen density. 

We estimate the electron density in two independent ways. First, \citet{compiegne_2007} used the S[III] 19 and 33 $\mu$m fine-structure lines to determine the electron density just ahead of the Horsehead (within $\sim$ 0.02 pc from the ionization front) to be $n_\mathrm{e}$ $\simeq$ 100-350 cm$^{-3}$. Second, the local density at the ionization front is estimated by calculating the amount of ionizing photons $J$ which land on the surface of the L1630 molecular cloud. Assuming a dust attenuation of $\mathrm{e}^{-\tau}$ = 0.5, \citet{abergel_2003} approximated this at $J$ = 0.8 $\times$ 10$^8$ cm$^{-2}$ s$^{-1}$. Dividing this by the isothermal sound speed {\bf $c_{\mathrm{s}}$ = ($kT$/$\mu m_{\mathrm{H}}$)$^{1/2}$} = 10 km s$^{-1}$ at $T_\mathrm{e}$ = 7500~K \citep{ferland_2003}, we calculate a density of $n_\mathrm{H}$ = $n_\mathrm{e}$ = $J$/$c_\mathrm{s}$ = 80 cm$^{-3}$, in close agreement with the value derived by \cite{compiegne_2007}.

The EM of the flow is derived using the calibrated KPNO H$\alpha$ intensity together with standard conversion factors taken from \cite{osterbrock_2006}. We measure EM = 1.8 $\times$ 10$^{4}$ cm$^{-6}$ pc at the ionization front. We can then pin down the path length $l$ of the L1630 molecular cloud at the cloud surface using the derived density $n_\mathrm{H}$. In this way, L1630 needs to extend 1 pc along the line of sight in order to reproduce the derived density. 

We are particularly interested in the gas density further downstream, where the ionized flow interacts with $\sigma$ Ori AB (see Fig. \ref{fig:mosaic_withzoom}). For this, we need to adopt a density law between the star and the molecular cloud. As discussed in Sec. \ref{sec:intro} and above, the density contrast between the tenuous \HII region and the dense molecular cloud L1630 sets up a champagne flow streaming into the \HII region \citep{tenorio_tagle_1979}. Strictly speaking, on a large scale the surface of the L1630 cloud will be convex, which will lead to a photo-evaporation flow following the nomenclature described in \citet{henney_2005}. However, as the radius of curvature of the ionization front is much larger than the distance to the ionizing source and as we are focussing on the material passing close to the star (which will originate from a small part of L1630), the analysis of \citet{henney_2005} suggest that we can approximate the structure of the flow as a champagne flow in a plane-parallel geometry. For this we will adopt the results for an isothermal shock tube as described in \citet{bedijn_1981} and \citet{tielens_2005}.

The initial velocity of the gas leaving the cloud surface is assumed to be comparable to the sound speed (i.e., a D-critical ionization front). The rarefaction wave moving into the ionized cloud gas sets up an exponential density gradient along which the gas is accelerated into the \HII region. The continuity and momentum equations can then be used to show that the gas follows a linear velocity law defined as

\begin{equation}
\label{eq:velocity}
dv_\mathrm{g} = c_\mathrm{s} \frac{d\rho_\mathrm{g}}{\rho_\mathrm{g}}.
\end{equation}

\noindent Here, $v_\mathrm{g}$ and $\rho_\mathrm{g}$ are the velocity and density of the flow. Rewriting this in terms of the number density $n$, Eq. \ref{eq:velocity} integrates to \citep{bedijn_1981}

\begin{equation}
\label{eq:density}
n(r) = \frac{J}{c_\mathrm{s}}\exp\left(\frac{r - R_\mathrm{s}}{c_\mathrm{s}t}\right),
\end{equation}

\noindent where $r$ is the distance from the star ($r$ $\leq$ $R_\mathrm{s}$); $R_\mathrm{s}$ is the Str\"{o}mgren radius \citep{stromgren_1939} (in our case: the distance between $\sigma$ Ori and the molecular cloud), and $t$ is the time which has passed since the material at the head of the flow had been ionized and started to move into IC 434. 

We adopt an expansion law to describe the evolution of the path length $l$ with distance from the molecular cloud. While the flow expands into the \HII region, we increase the path length linearly. When the flow has reached $\sigma$\,Ori AB, we assume that the scale size of the emitting gas along the line of sight is equal to the traversed distance, i.e., the projected distance between L1630 and $\sigma$\,Ori AB of $d$ = 3.2 pc at a distance of 334 pc. The ionized flow then shows a smooth drop in hydrogen density, ranging from $n_\mathrm{H}$ = 100 cm$^{-3}$ at the cloud surface to $n_\mathrm{H}$ = 10 cm$^{-3}$ near $\sigma$ Ori~AB. The result is shown in Fig. \ref{fig:halphacut}. We find that, due to the pressure gradient, the gas near $\sigma$ Ori~AB is accelerated to 35 km s$^{-1}$ which is 3.5 times the local sound speed $c_\mathrm{s}$. The timescale for the ionized gas to reach the star is then roughly 1.5 $\times$ 10$^{5}$ yr. Taking into account the space velocity of $\sigma$ Ori\,AB, we calculate a maximum relative velocity between the star and the ionized flow of 50 km s$^{-1}$. In summary, Fig. \ref{fig:halphacut} demonstrates that the density law is well described by an exponential density law with a scale height of 1 pc at the cloud surface. As the density at the ionization front is well determined from the intensity ratio of the S[III] fine structure lines, the only assumption that enters into this comparison is the linear expansion of the flow along the line of sight from 1 pc at the ionization front to $\sim$3 pc at $\sigma$ Ori.

\section{Bow waves and dust waves}\label{sec:structure}

Figure \ref{fig:mosaic_withzoom} reveals an increase in emission at mid-IR wavelengths near $\sigma$ Ori AB, peaking at a distance of $d$ = 0.1 pc ahead of the star. Given that $\sigma$ Ori AB is classified as a weak-wind candidate, we propose that the extended emission surrounding $\sigma$ Ori AB represents the first detection of a radiation-pressure driven structure around a massive star moving through the ISM \citep{van_buren_1988}. The radiation pressure of $\sigma$ Ori AB acts on the dust, stalling it at a distance ahead of the star where the ram pressure of the ISM material is balanced. We adjust the nomenclature as described by \citet{van_buren_1988} and distinguish between a {\em dust wave}, where dust is stopped and decouples from the gas, and a {\em bow wave}, where dust is stopped and gas stays coupled. This section will elaborate on information extracted from the IR observations, while Sec. \ref{sec:flow} will deal with the physics of bow waves and dust waves.

\subsection{Dust distribution}

\begin{figure*}
 \captionsetup[subfigure]{labelformat=empty}  
  \subfigure{\includegraphics[width=8.5cm]{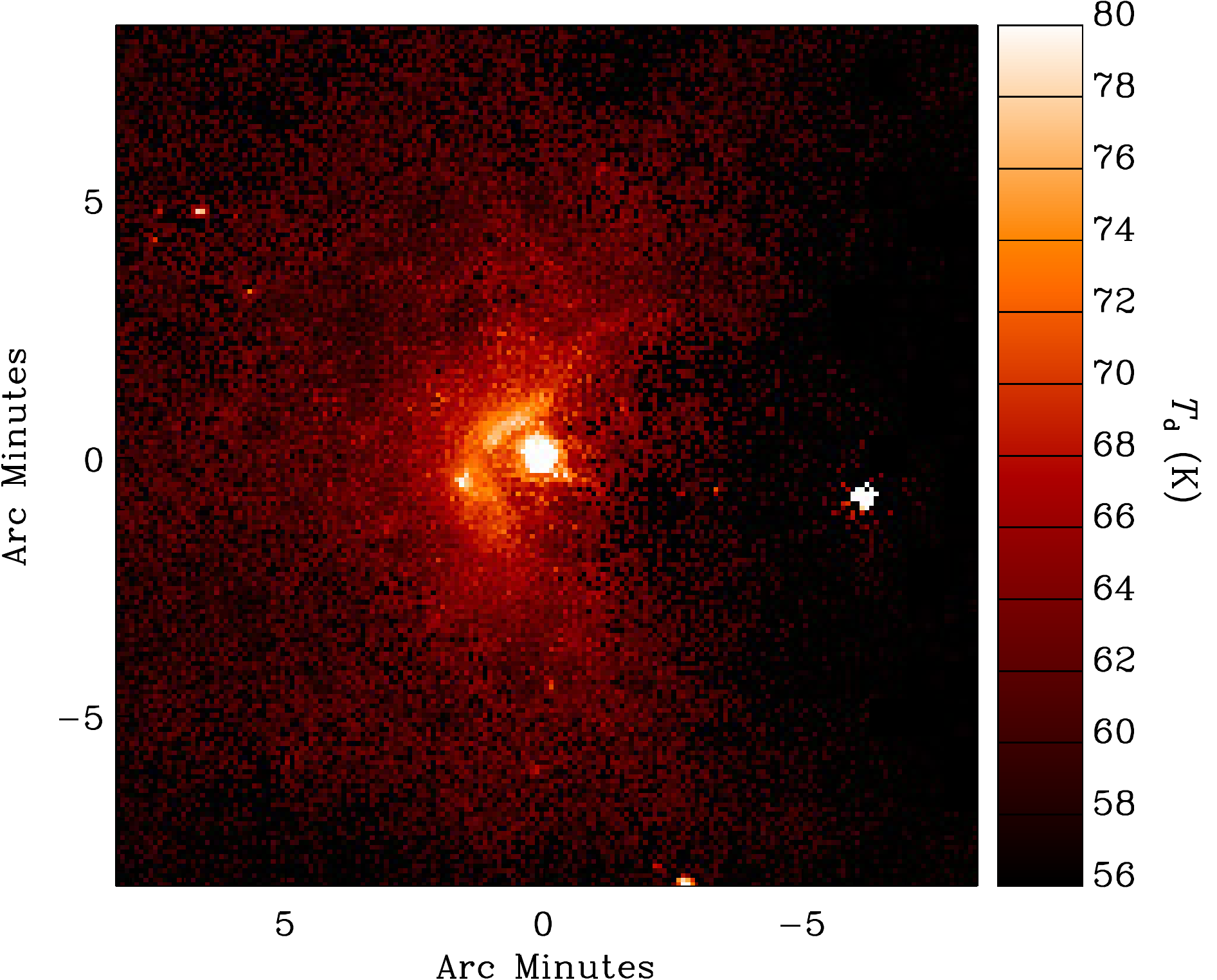}\label{fig1:temperature}}
  \qquad
  \subfigure{\includegraphics[width=9cm]{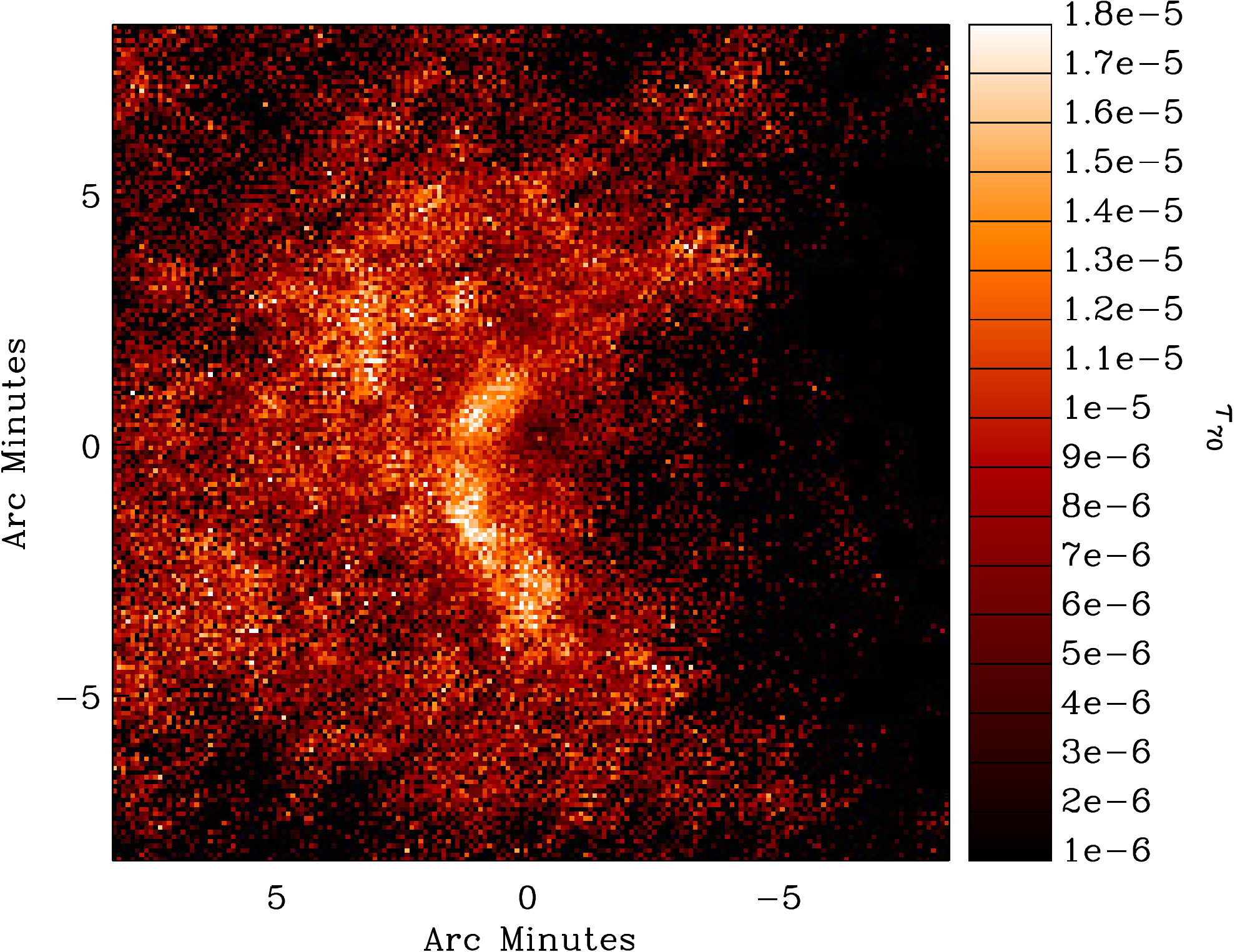}\label{fig1:opticaldepth}}
\caption{{\em Left}: MIPS 24 $\mu$m versus 70 $\mu$m PACS dust color temperature map. The scale size is 16.5$\arcmin$ $\times$ 16.5$\arcmin$. Only pixels with values $\textgreater$ 3$\sigma$ in the PACS 70 $\mu$m image are plotted. North is up, east is to the left. {\em Right}: 70 $\mu$m optical depth map.}
  \label{fig1:tempopt}
\end{figure*}

The dust color temperature is derived from the 24 $\mu$m and 70 $\mu$m intensities after convolving the PACS 70 $\mu$m image to the MIPS 24 $\mu$m beam using the convolution kernels described in \citet{aniano_2011}. According to the Herschel PACS Instrument Calibration Centre (ICC), photometric measurements between PACS and MIPS are consistent within 17\%\footnote{PICC-NHSC-TN-029: https://nhscdmz2.ipac.caltech.edu/pacs/docs/\\Photometer/PICC-NHSC-TN-029.pdf}. The dependence of dust emissivity on wavelength, $\beta$, is of little importance in our study as we are extrapolating towards long wavelengths where the intensity is low. In this work, we fix the emissivity at $\beta$ = 1.8. Following \citet{hildebrand_1983}, the dust optical depth is estimated by

\begin{equation}
\label{eq:opticaldepth}
\tau_{\nu} = \frac{I_\nu}{B_{\nu,T_\mathrm{d}}},
\end{equation} 

\noindent where $I_\nu$ is the surface brightness and $B_\mathrm{\nu}$ is the Planck function for dust temperature $T_\mathrm{d}$. We evaluate Eq. \ref{eq:opticaldepth} at 70 $\mu$m. The dust color temperature map and the optical depth map at 70 $\mu$m are plotted in Fig. \ref{fig1:tempopt}. Dust  temperatures range from 50$\pm$10 K to 75$\pm$14 K near $\sigma$ Ori AB while the optical depth $\tau_\mathrm{70}$ ahead of the star increases by a factor of 2-3 compared to the region behind $\sigma$\,Ori AB.

\subsection{Dust mass}

The spectral energy distribution (SED) of the emission is extracted using an aperture which encapsulated the IR emission while minimizing stellar contamination by $\sigma$ Ori AB. This allows us to study the average dust properties within the aperture. Subsequently, an average of several background positions absent of obvious emission is subtracted. A two-component modified blackbody function is then fitted to the IRAC 8.0 $\mu$m, WISE 12 $\mu$m, MIPS 24 $\mu$m and PACS 70 $\mu$m integrated intensities. We minimize the temperature variation across the aperture by focussing on the bright emission only. The IRAC 3.4, 4.6 and 5.8 $\mu$m images are heavily affected by diffraction patterns from $\sigma$ Ori AB and could therefore not be used to extract reliable flux values. The two components are believed to represent different dust populations. The emission component peaking near 45 $\mu$m, radiating at a temperature $T$ = 68 K, is thought to originate from large dust grains or Big Grains (BGs), which are in thermal equilibrium with the radiation field. The other component emitting at shorter wavelengths, radiating at $T$ = 197 K, probably originates from Very Small Grains (VSGs). In typical ISM conditions, the VSGs are stochastically heated. Here, it is safe to assume that the intensity of the radiation field is high enough for the VSGs to be in thermal equilibrium and can therefore be fitted by a modified blackbody function.

\begin{figure}         
\includegraphics[width=9cm]{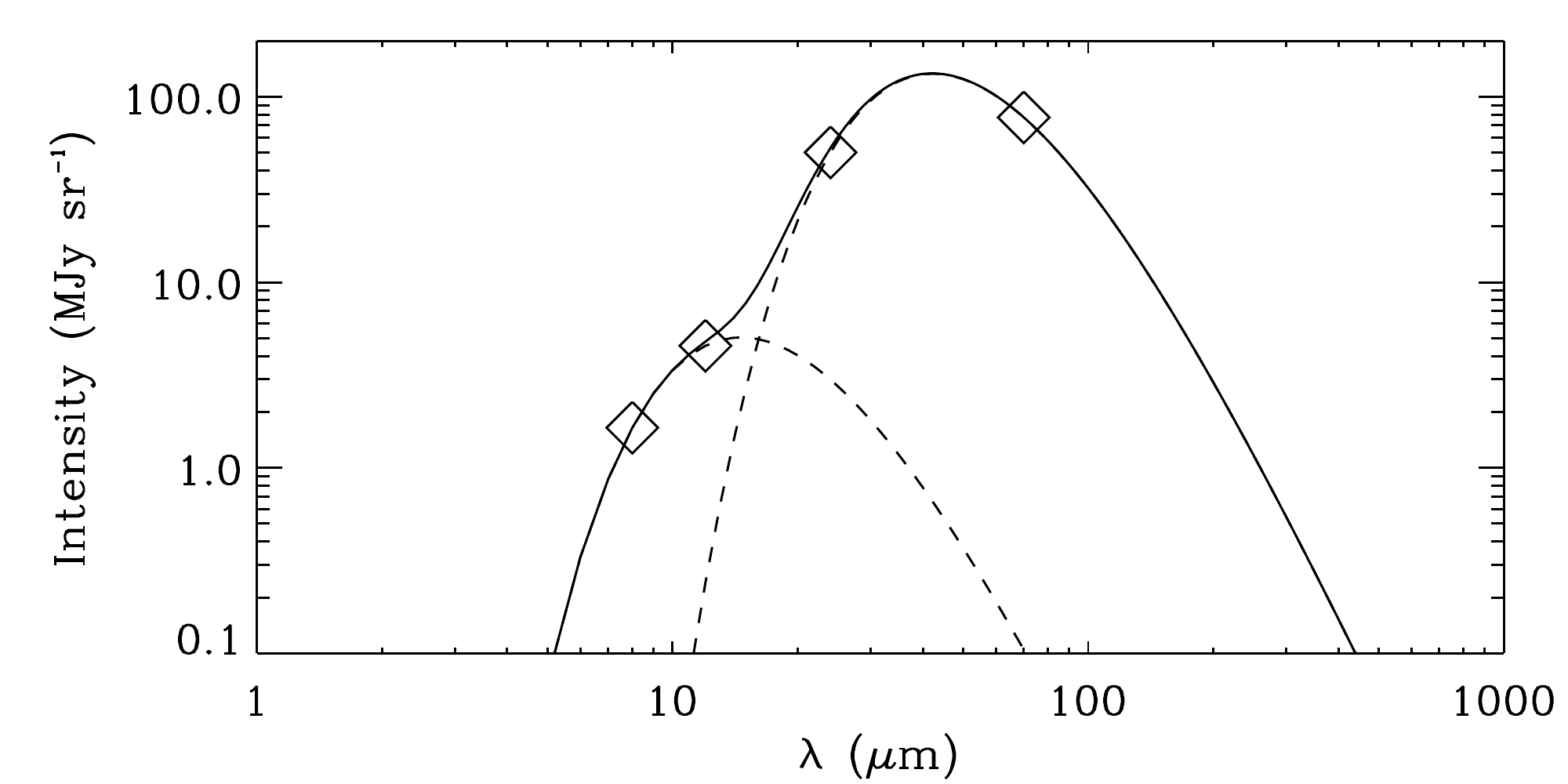}
 \label{fig:apertureandsed}
 \caption{A two-component modified black body is fitted to SED of the extended IR emission around $\sigma$ Ori AB.}
\end{figure}

\subsubsection{UV derived mass}

The total surface brightness $I_\mathrm{tot}$ is found by integrating the SED over frequency, yielding $I_\mathrm{tot}$ = 7.6 $\times$ 10$^{-3}$ erg s$^{-1}$ cm$^{-2}$ sr$^{-1}$. The total luminosity $L_\mathrm{IR}$ is then $L_\mathrm{IR} = 4 \pi I_\mathrm{tot} S$, where $S$ is the emitting surface area of the IR structure. With $d$ = 334$^{+25}_{-22}$ pc and an aperture which subtends 1.3 $\times$ 10$^4$ square arcseconds on the sky, we have $S$ =  3.4$\pm$0.5 $\times$ 10$^{35}$ cm$^2$ and thus $L_\mathrm{IR}$ = 8.3$\pm$1.6 $L_\odot$. The stellar luminosity is taken from \citet{lee_1968} and equals $L_{\ast}$ = 0.6 $\times$ 10$^5$ $L_\odot$. The energy absorbed by the dust grains through UV photons is subsequently re-emitted in the IR.  Therefore, a good estimate for the optical depth at UV wavelengths is given by the ratio  $L_\mathrm{IR}$/$L_{\ast}$ = $\tau_\mathrm{UV}$ =  1.4$\pm$0.3 $\times$ 10$^{-4}$. From the optical depth at UV wavelengths, we can estimate the total dust mass $M_\mathrm{d,UV}$ by assuming a grain opacity at UV wavelengths $\kappa_\mathrm{UV}$ and by multiplying with the surface area of a shell located at $r$ = 0.1 pc:

\begin{equation}
\label{eq:dustmass_uv}
M_\mathrm{d,UV} = 4 \pi  r^2 \frac{\tau_{\mathrm{UV}}}{\kappa_{\mathrm{UV}}}.
\end{equation}

\noindent We take $\kappa_\mathrm{UV}$ = 1 $\times$ 10$^{4}$ cm$^{2}$ g$^{-1}$ \citep{weingartner_2001}. In this way, we calculate a total dust mass of $M_\mathrm{d,UV}$ = 8.4$\pm$1.6 $\times$ 10$^{-6}$ $M_\odot$. 

\subsubsection{IR derived mass}

It is possible to obtain an estimate of the total dust mass through the IR observations. $M_\mathrm{d,IR}$ is directly proportional to the detected far-IR emission:

\begin{equation}
\label{eq:totalmass}
M_\mathrm{d,IR} = \frac{\tau_{\mathrm{\nu}}}{\kappa_\mathrm{\nu}} S.
\end{equation}

\noindent Here, $\kappa_\mathrm{\nu}$ = 60 cm$^2$ g$^{-1}$ at 70 $\mu$m \citep{weingartner_2001} and the average dust optical depth within the aperture is  $\tau_{70}$ = 1.3$\pm$0.2 $\times$ 10$^{-5}$. We derive a total dust mass $M_\mathrm{d,IR}$ = 3.7$\pm$0.7 $\times$ 10$^{-5}$ $M_\odot$. 

\subsubsection{Summary on the observations}

The total dust mass is approximated both through energy absorbed in the UV ($M_\mathrm{d,UV}$) or directly through the IR 70 $\mu$m emission ($M_\mathrm{d,IR}$). Our calculations show a discrepancy between both derived dust masses, i.e., $M_\mathrm{d,IR}$/$M_\mathrm{d,UV}$ = 4$\pm$0.8. This discrepancy may simply reflect that the dust opacities at UV and far-IR wavelengths for dust in \HII regions is very different from that calculated for (average) interstellar dust. In particular, \citet{weingartner_2001} use a standard Mathis-Rumpl-Nordsieck (MRN) size distribution \citep{mathis_1977}, although the size distribution can be heavily affected during coagulation in the molecular cloud. A size distribution with relatively more large dust grains would reduce the opacity at UV wavelengths. A discrepancy between far-IR and UV derived dust abundances has been noted by \citet{salgado_2012} for dust in \HII regions. Considering the discrepancy between values derived in both methods, we adopt a total dust mass of $M_\mathrm{d}$ = 2.3$\pm$1.5 $\times$ 10$^{-5}$ $M_\odot$, which is the mean value of both earlier derived values (with the 1$\sigma$ uncertainty). The total dust-to-gas mass fraction can then be approximated by comparing $M_\mathrm{d}$ with the amount of gas $M_\mathrm{g}$ contained in the same volume. We adopt a geometry of a spherical shell with thickness $dr$ $\approx$ 25$\arcsec$ = 1.3 $\times$ 10$^{17}$ cm. The S[III] spectra show that there is no clear detection of an enhanced density structure in the gas coinciding with the observed dust arc (although high-resolution imaging is needed to conclude on the (non-)existence of a gaseous structure around $\sigma$ Ori AB; see the discussion in Sec. \ref{sec:bowshock}). Here, we opted to extrapolate the large scale density distribution measured from the KPNO H$\alpha$ data. The gas density at the location of the dust arc is then about 10 cm$^{-3}$ (Sec. \ref{sec:emissionmeasure}). With $n_\mathrm{H}$ = 10 cm$^{-3}$, we calculate a total gas mass of $M_\mathrm{g}$ = 6.9$\pm$1.3 $\times$ 10$^{-5}$ $M_\odot$. The total dust-to-gas mass ratio is then $M_\mathrm{d}$/$M_\mathrm{g}$ = 0.29$\pm$0.20, significantly higher than the standard ratio in the ISM ($\sim$ 0.01). The derived dust-to-gas ratio implies that the dust number density has increased; we attribute this to radiation pressure acting on the dust grains, creating a {\em dust wave} ahead of $\sigma$ Ori AB.

\section{Physics of bow waves and dust waves}\label{sec:flow}

The dust wave around $\sigma$ Ori\,AB is continuously fed by the photo-evaporation flow emanating from the L1630 molecular cloud. This flow is considered to be a two-component fluid of dust and gas particles, each of the components containing its own density, temperature and velocity structure. 
We follow the flow from L1630 to $\sigma$ Ori\,AB. As the gas is being accelerated and approaches the star, it will drag along the dust grains through collisions. Inside the flow, dust grains will absorb photon energy and photon momentum from the ionizing star: while the energy is radiated away, the absorbed momentum causes the dust to be decelerated. We do not consider interaction with the stellar wind; it is assumed that the momentum flux of the wind is negligible compared to the radiation pressure. The ionized gas component of the flow has a negligible cross section for photon absorption, but exchanges momentum with the dust through drag. Through collisions of gas atoms with a dust grain, a significant amount of momentum from the dust can be transferred to the gas. This momentum is then distributed over all gas particles by internal collisions. Depending on the magnitude and efficiency of the momentum transfer, the gas and the dust component can stay coupled in the flow. The dust wave around $\sigma$ Ori\,AB is caused by a balance between radiation pressure and the drag force. Below, we present a quantitative model describing the physics of this interaction.

\subsection{Dust component}

We write the equation of motion for dust as \citep{tielens_1983}

\begin{equation}
\label{eq:eqmotiondust}
m_\mathrm{d}\frac{dv_\mathrm{d}}{dt} = -\frac{\sigma_\mathrm{d} \bar{Q}_\mathrm{rp} L_{\star}}{4 \pi cr^2} + F_\mathrm{drag} + F_\mathrm{Lorentz},
\end{equation}

\noindent where $m_\mathrm{d}$ is the mass of the dust grain; $r$ is the distance from the star; $L_{\star}$ and $c$ are the luminosity of the star and the speed of light; and $\sigma_{\mathrm{d}}$ and $\bar{Q}_{\mathrm{rp}}$ are the geometrical cross section and the flux weighted mean radiation pressure efficiency of the grains. The first term on the right hand side of Eq. \ref{eq:eqmotiondust} describes the radiation pressure force $F_\mathrm{rad}$. The term $F_\mathrm{Lorentz}$ is the Lorentz force and is described below. $F_\mathrm{drag}$ is the drag force due to interactions of the dust with atomic or ionic species $i$ and equals \citep{draine_salpeter_1979}:

\begin{equation}
\label{eq:dragforce}
F_\mathrm{drag} = 2\pi a^2kT n_i \left(G_1(s_i) + z_i^2 \phi^2 ln(\Lambda/z_i) G_2(s_i)\right).
\end{equation}

\noindent Here, $s_i = \sqrt{m_i v_\mathrm{drift}^2/2kT}$, $m_i$ and $z_i$ are the mass and charge of the interacting atoms or ions; $a$ is the grain radius and $v_\mathrm{drift}$ is the relative velocity between the gas and the dust called the drift velocity. The drag force includes both the direct drag through collisions and the plasma drag through long range Coulomb interactions with ionic species and electrons. The electrostatic grain potential is defined by $\phi = Z_\mathrm{d} e^2 / kT$, where $Z_\mathrm{d}$ is the grain charge and $e$ is the elementary charge. $\Lambda$ represents the Coulomb factor \citep{spitzer_1978}:

\begin{equation}
\label{eq:coulomb_factor}
\Lambda = \frac{3}{2ae |\phi|}\left(\frac{kT}{\pi n_\mathrm{e}}\right).
\end{equation}
 
\noindent The functions $G_\mathrm{1}$ and $G_\mathrm{2}$ are approximated by \citep{baines_1965,draine_salpeter_1979}

\begin{equation}
\label{eq:g1}
G_\mathrm{1} (s) \approx \frac{8s}{3 \sqrt{\pi} } \left(1 + \frac{9\pi}{64}s^2\right)^{1/2}  
\end{equation}

\noindent and

\begin{equation}
\label{eq:g2}
G_\mathrm{2} (s) \approx s\left( \frac{_3}{^4} \pi^{1/2} + s^3\right)^{-1}.
\end{equation}

\noindent Grains with a velocity component perpendicular to the local magnetic field \vec{B} will gyrate around the field lines due to the Lorentz force $F_\mathrm{Lorentz}$:

\begin{equation}
\label{eq:lorentzforce}
F_\mathrm{Lorentz} = m_\mathrm{d} \vec{v} \times \vec{\omega}_B.
\end{equation}

\noindent The angular velocity $\vec{\omega_B}$ is given by

\begin{equation}
\label{eq:angular_velocity}
\vec{\omega_\mathrm{B}} = \frac{z_\mathrm{d} e}{m_\mathrm{d} c} \vec{B}.
\end{equation}

\noindent The drift velocity between the gas and the dust component can lead to the ejection of atoms from the surface of the dust into the gas phase. Grain-grain collisions will be negligible in the flow due to the low number density of grains compared to the gas. At low energies and for light projectiles, sputtering is caused by the reflection of an ion in a deeper layer of the grain. After the ion gets reflected, it knocks a surface atom into the gas phase. The rate of decrease in grain size is given by \citep{tielens_1994}

\begin{equation}
\label{eq:rateofchange2}
\frac{da}{dt} = \frac{m_\mathrm{sp}}{2\rho_\mathrm{s}}v_\mathrm{drift} n_i Y_i,
\end{equation}

\noindent where $m_\mathrm{sp}$ and $\rho_\mathrm{s}$ are the mass of the sputtered atoms and the specific density of the grain material; and $n_\mathrm{i}$ and $Y_\mathrm{i}$ are the number density of the gas particles and the sputtering yield. 

\subsection{Gas component}

The gas momentum equation does not contain the radiative deceleration term as the ionized hydrogen has a negligible cross section for photon absorption. We neglect gravitational attraction and therefore the equation of motion of gas consists of a balance between the pressure factor (Eq. \ref{eq:velocity}) and the momentum transfer through interactions with dust grains:

\begin{equation}
\label{eq:eqmotiongas}
\frac{dv_\mathrm{g}}{dt} = v_\mathrm{g}\frac{c_\mathrm{s}}{\rho_\mathrm{g}}\frac{d\rho}{dr} + \frac{n_\mathrm{d}}{\rho_\mathrm{g}} F_\mathrm{drag}.
\end{equation}

\noindent The term on the right hand side represents the amount of momentum transferred per second to a unit mass of gas. The dust number density $n_\mathrm{d}$ is taken from the MRN size distribution and the gas density $\rho_\mathrm{g}$ is adopted from Eq. \ref{eq:density}. The ratio $n_\mathrm{d}$/$\rho_\mathrm{g}$ is not constant, but will depend on the relative velocity of the gas and dust through the continuity equation. 

\subsection{Modeling the interaction between the flow and $\sigma$ Ori AB}

To determine the velocity structure of the photo-evaporation flow, we integrate the coupled differential equations \ref{eq:eqmotiondust}, \ref{eq:rateofchange2} and \ref{eq:eqmotiongas} simultaneously using a fourth-order Runge-Kutta method in three dimensions. As we neglect grain-grain collisions, we could replace Eq. \ref{eq:eqmotiondust} with a set of equations for each grain size. The drag force in Eq. \ref{eq:eqmotiongas} should then be rewritten to account for the size distribution. For simplicity, we assume a single dust size distribution, but we will investigate the influence of grain size by solving the equations for several values of $a$. First, we consider a flow without grain charge and show that this reproduces the observations. Afterwards, we will discuss the inclusion of the Coulomb drag term described in Eq. \ref{eq:dragforce} and the Lorentz force.

\begin{figure*}
  \centering
  \subfigure[Dust trajectories for a 3000 \AA\ silicate grain trajectories for different values of  impact parameter $b$.]{\includegraphics[width=8cm]{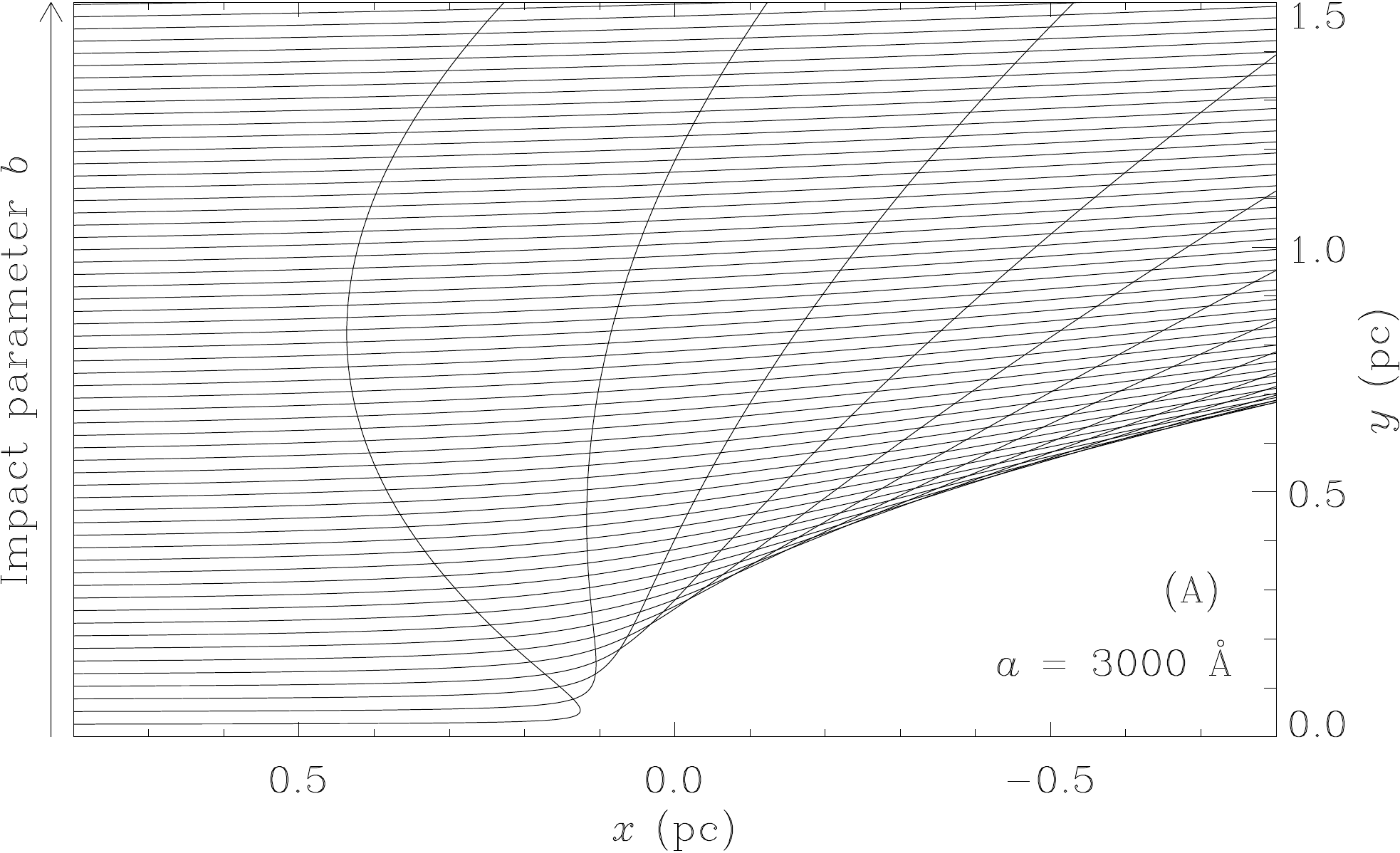}\label{fig1:300}}
\qquad
  \subfigure[Same, but for a different value of $\kappa_\mathrm{rp}$, which represents a 300 \AA\ silicate grain.]{\includegraphics[width=8cm]{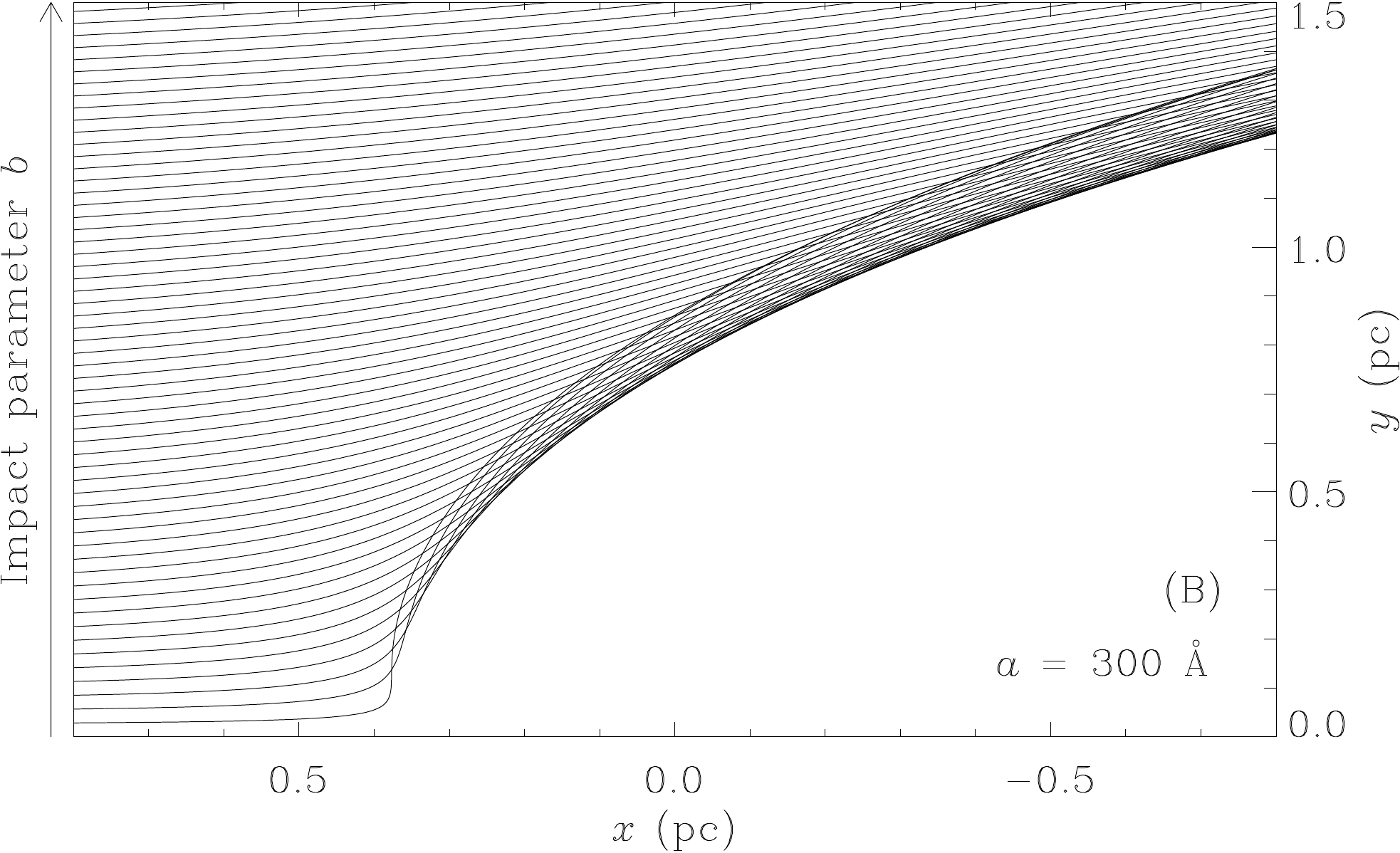}\label{fig1:2000}}
\caption{Two dimensional plots of dust grain streamlines, evaluated for different grain sizes $a$. The x-axis represents the projected distance from L1630 to $\sigma$ Ori, while the $y$-axis is the direction perpendicular to the line connecting L1630 and $\sigma$ Ori AB, in the plane of the sky.}
  \label{fig:dust_trajectories}
\end{figure*}

In our model, the $x$-axis is directed along the flow, which represents the projected distance from L1630 to $\sigma$ Ori AB. The star is placed at the origin. The $y$-axis and $z$-axis represent the directions perpendicular to the flow, i.e., the direction in the plane of sky and the direction along the line of sight, respectively. The flow starts at the molecular cloud, positioned at $x$ = 3.2 pc. We define the impact parameter $b$ as the distance along the y-axis (see Fig. \ref{fig:dust_trajectories}) at the start of the flow. A dust grain which approaches the star head-on ($b$ = 0) will lose momentum by absorbing photons and be stopped at the point where the drag force $F_\mathrm{drag}$ balances the radiation pressure force $F_\mathrm{rad}$. More specifically, an incoming dust grain will tend to overshoot its stopping distance, reaching a minimum distance to the star $r_\mathrm{min}$ depending on its momentum. Subsequently, it is pushed back by the radiation pressure force until an equilibrium radius $r_\mathrm{eq}$ is reached where $|F_\mathrm{rad}$/$F_\mathrm{drag}|$ equals unity. This results in a damped oscillation around $r_\mathrm{eq}$. Particles with $b$ $\textgreater$ 0 will have similar trajectories, but will gain momentum in the $y$ direction and therefore be pushed past the star in a shell-like structure. We define the radiation pressure opacity $\kappa_\mathrm{rp}$, which determines the exact trajectory of a particle

\begin{equation}
\label{eq:control}
 \kappa_\mathrm{rp} = \sigma_\mathrm{d} \bar{Q}_\mathrm{rp} / m_\mathrm{d}.
\end{equation}

\noindent We evaluate the model for two different values of $\kappa_\mathrm{rp}$. For small grains, $\kappa_\mathrm{rp}$ is higher because the ratio of surface area over volume increases with smaller $a$. We have chosen the values of $\kappa_\mathrm{rp}$ in a way to clearly demonstrate the influence of grain size. The momentum of the incoming particle is determined by the grain size and the specific density $\rho_\mathrm{s}$. In both models we choose $\rho_\mathrm{s}$ = 3.5 g cm$^{-3}$, intermediate between the specific density of crystalline forsterite (3.21 g cm$^{-3}$) and fayalite (4.39 g cm$^{-3}$). For comparison, ideal graphite has a specific density of 2.24 g cm$^{-3}$. We set $\bar{Q}_\mathrm{rp}$ = 1 throughout the flow in both models. In reality, $\bar{Q}_{\mathrm{rp}}$ is dependent on $r$ because it is averaged over the photon spectrum. This spectrum is altered due to attenuation of dust inside the \HII region. The total dust optical depth is estimated at e$^{-\tau}$ = 0.5 \citep{abergel_2003} and the effect of the dust wave on the UV radiation field is negligible (see Sec. \ref{sec:structure}). The radiation pressure becomes important close to the star where the dust attenuation is low, which makes $\bar{Q}_{\mathrm{rp}}$ independent of $r$ plausible. The initial dust number densities per H atom ($n_\mathrm{d}/n_\mathrm{H}$) are estimated using the MRN size distribution \citep{mathis_1977} where we consider both the contribution of graphite and silicate grains. In this way, small grains have an abundance of 5.7 $\times$ 10$^{-11}$ H-atom$^{-1}$ for a bin size ranging from 100 \AA\ to 500 \AA. For the big grains, we use a bin size ranging from 1000 \AA\ to 5000 \AA. These grains have an abundance of 1.8 $\times$ 10$^{-13}$ H-atom$^{-1}$. The gas component follows the same density and velocity law as derived in Sec. \ref{sec:emissionmeasure}. We calculate {\em a posteriori} the amount of momentum which the gas has acquired through collisions with the dust.

\begin{table}
\caption{Parameters used in calculating the two different models shown in Fig. \ref{fig:dust_trajectories}. $a$ is the radius of the grain and $\rho_\mathrm{s}$ is the specific density of the grain material. These combinations of parameters represent a fiducial grain and do not necessarily represent grains taken from existing dust models.}
\label{tab:model}
\begin{tabular}{llllll|l|}\hline\hline
& Model A & Model B & Units \\ \hline 
$\kappa_\mathrm{rp}$ & 7 $\times$ 10$^{3}$ & 7 $\times$ 10$^{4}$ & cm$^2$ g$^{-1}$ \\
$a$ & 3000 & 300 & \textup{\AA} \\
$n_\mathrm{d}$/$n_\mathrm{g}$ & 1.8 $\times$ 10$^{-13}$ & 1.7 $\times$ 10$^{-11}$ & \\ \hline
\end{tabular}
\end{table}

\section{Results}\label{sec:results}

In Fig. \ref{fig:dust_trajectories} we plot trajectories along which a specific dust grain will travel. Model A represents a 3000 \AA\ grain ($\kappa_\mathrm{rp}$ =  7 $\times$ 10$^{3}$ cm$^2$ g$^{-1}$), while model B represents a 300 \AA\ grain ($\kappa_\mathrm{rp}$ = 7 $\times$ 10$^{4}$ cm$^2$ g$^{-1}$) with identical values for $\rho_\mathrm{s}$ (= 3.5 g cm$^{-3}$) and $\bar{Q}_\mathrm{rp}$ (= 1). The effect of dust particle size is clearly seen. For larger values of $\kappa_\mathrm{rp}$, radiation pressure is able to alter the direction of the grain further away from the star, even for trajectories with small initial $b$. While the distance which the particle overshoots is far less compared to model A, the stopping distance of the grain increases with high $\kappa_\mathrm{rp}$. We calculate that a dust grain with  $\kappa_\mathrm{rp}$ = 6.7 $\times$ 10$^{3}$ cm$^2$ g$^{-1}$ (3300 \AA\,) will be stopped at $d$ = 0.1 pc, which coincides with the peak emission of the dust wave at 24 $\mu$m.

\begin{figure*}
  \centering
  \subfigure[The velocity, density increase and the amount of momentum transfer, evaluated for a 300 \AA\ dust grain with impact parameter $b$ = 0.01.]{\includegraphics[width=8cm]{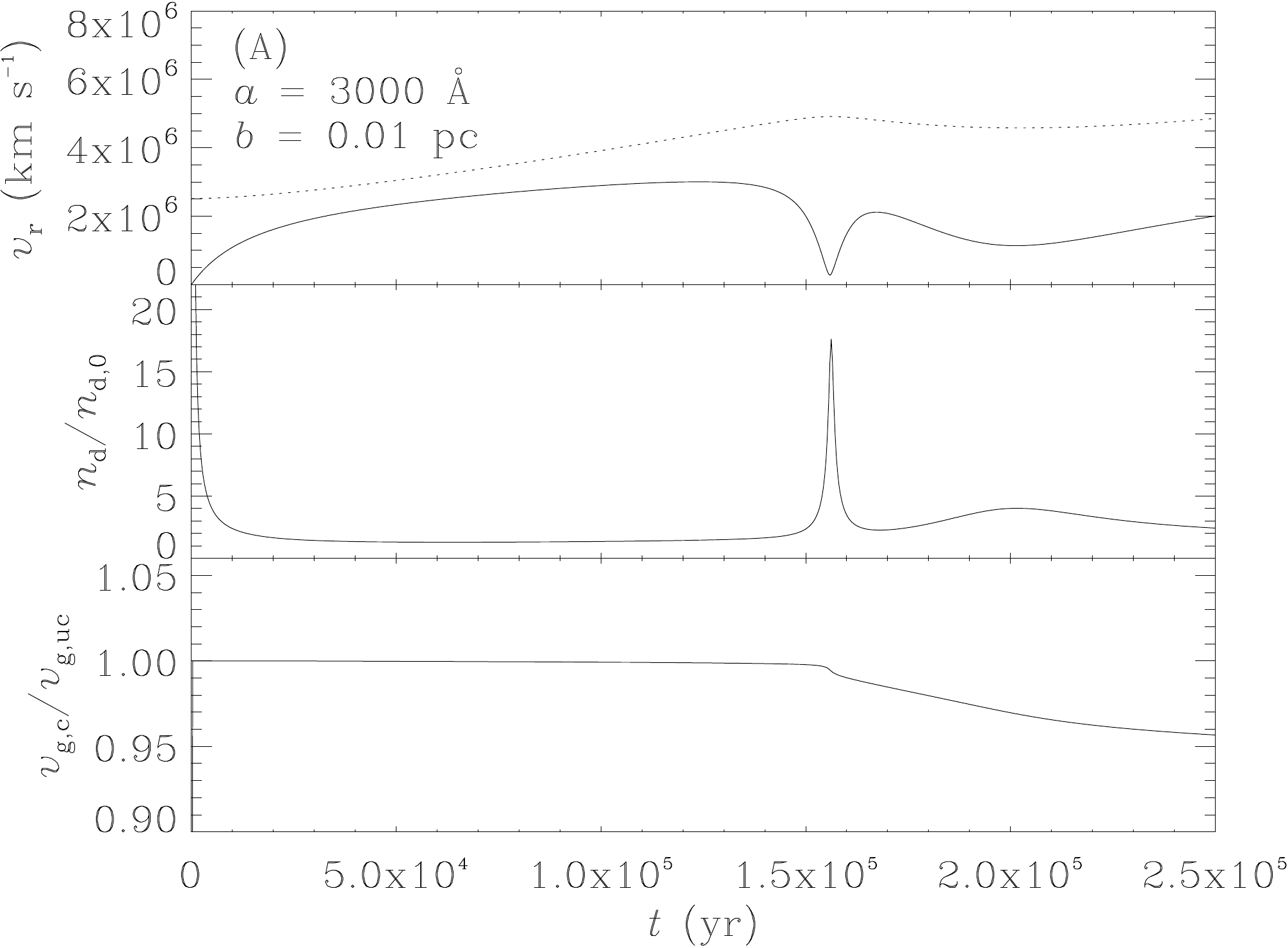}\label{fig1:300}}
  \subfigure[Same figures, but for a dust grain with a 300 \AA\ radius.]{\includegraphics[width=8cm]{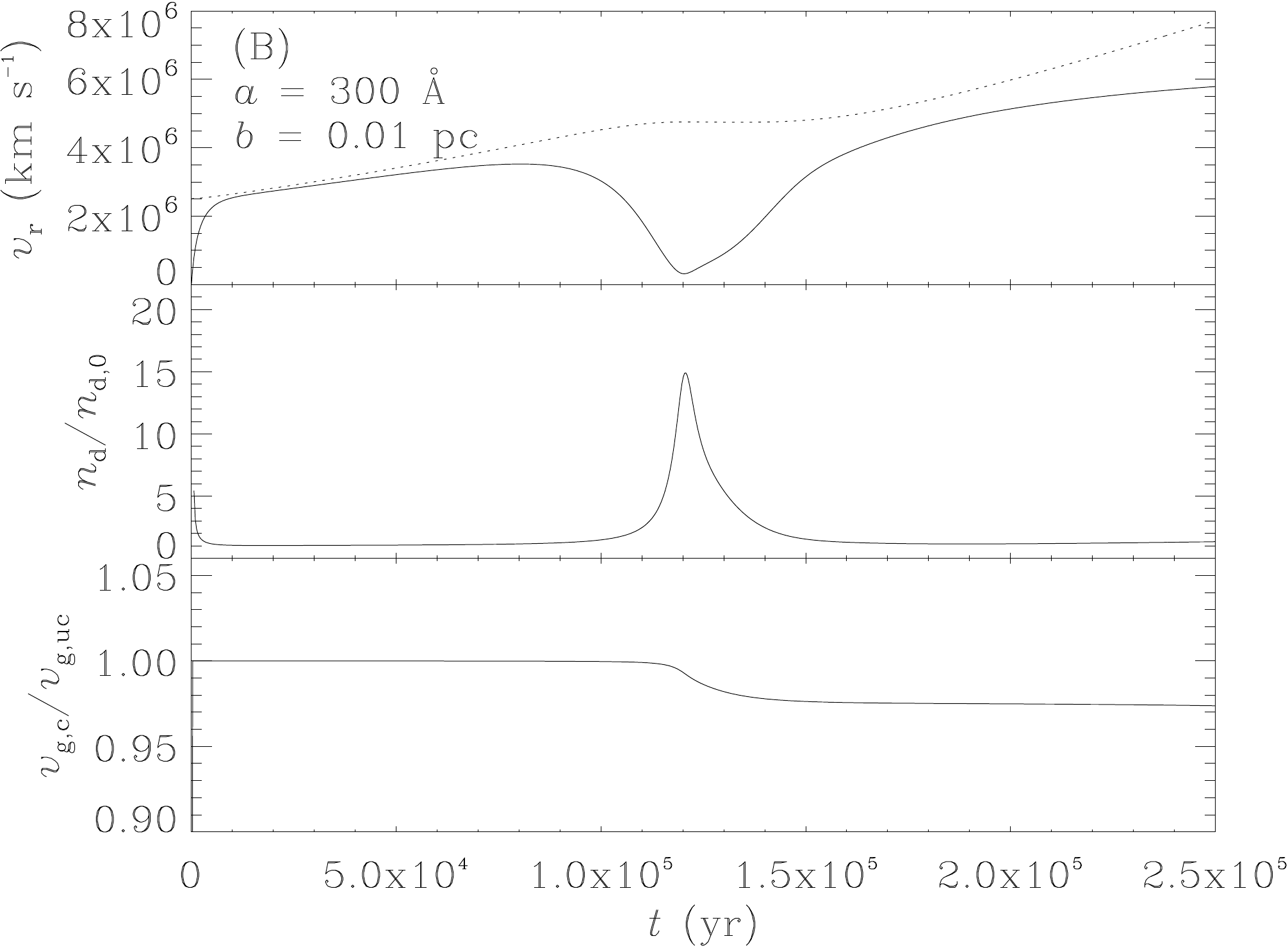}\label{fig1:2000}}
\caption{{\em Upper panels}: The radial velocity $v_\mathrm{r}$ of a dust grain (solid) and a gas particle (dotted, evaluated at the position of the dust grain in the frame of the star) as a function of time. At time $t$ = 0, the gas is ionized and enters the \HII region. {\em Middle panels}: The relative density increase in dust along the trajectory. This value approaches unity when the gas and dust are coupled. {\em Lower panels}: The velocity ratio of a gas particle $v_\mathrm{g,c}$, after momentum transfer with the dust, to an uncoupled gas particle $v_\mathrm{g,uc}$, which reflects the amount of momentum transferred from the dust to the gas.}
  \label{fig:dust_gas_trajectories}
\end{figure*}

Figure \ref{fig:dust_gas_trajectories} shows results for one specific trajectory with initial impact parameter $b$ = 0.01 pc. Our calculations show that a 300 \AA\ grain will be accelerated on a short timescale, reaching the velocity of the gas at $t$ =  2.5 $\times$ 10$^{3}$ yr. Momentum is lost gradually and $r_\mathrm{min}$ is reached after 1.5 $\times$ 10$^{5}$ yr, where the relative increase in number density peaks at 17 times the initital number density ($n_\mathrm{d,0}$). Figure \ref{fig:dust_gas_trajectories} also shows the ratio between the velocity of the gas with ($v_\mathrm{g,c}$) and without ($v_\mathrm{g,uc}$) momentum transfer with the dust. The gas will flow at a velocity of 0.97$v_\mathrm{g,uc}$ past the star and will therefore lose 3\% of its initial momentum.  Fig. \ref{fig:dust_gas_trajectories} shows that the large grains (3000 \AA\,) will be decelerated by the radiation pressure force before reaching the gas velocity. Even though the 3000 \AA\ grains have lower velocities relative to the star compared to 300 \AA\ grains, they have significantly more momentum and will therefore approach the star more closely. The time spent at $r_\mathrm{min}$ is smaller because the radiation pressure force is far greater close to the star. However, we  calculate that the total momentum transferred to the gas is small: gas flows at 0.96$v_\mathrm{g,uc}$ past the star. 

\begin{figure}
\includegraphics[width=9cm]{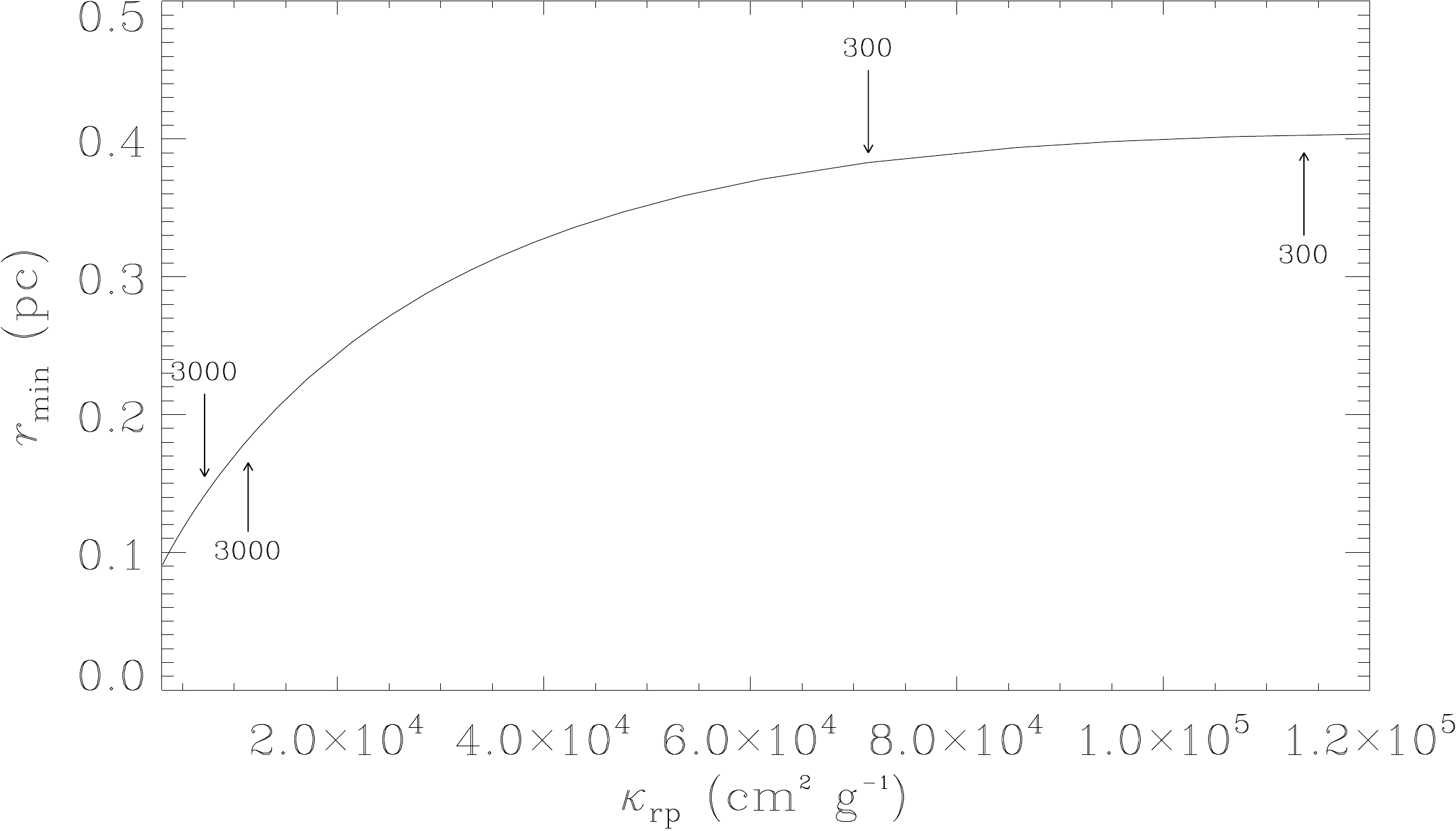}
\caption{The point of closest approach to the star $r_\mathrm{min}$  versus radiation pressure opacity $\kappa_\mathrm{rp}$ for  trajectories with $b$ = 0, assuming $\bar{Q}_\mathrm{rp}$ = 1 independent of grain size $a$. Labeled above the curve are typical grain sizes in angstroms for silicates. Below the curve are the corresponding grain size but for graphite grains.}
\label{fig:stopping_radius}
\end{figure}

The momentum of a grain depends on its size and composition, which will ultimately determine the trajectory of the particle. Larger grains will tend to move closer to the star compared to smaller grains. Consequently, dust will be stratified according to their size and specific properties. Figure \ref{fig:stopping_radius} plots the minimum radius $r_\mathrm{min}$ versus the radiation pressure opacity  $\kappa_\mathrm{rp}$. When comparing silicate and graphite grains of similar sizes, graphite particles are stopped further out from the star because of their low specific density.

The threshold energy for sputtering of hydrogen atoms from an amorphous carbon and silicate surface is $E$ = 0.5$m_\mathrm{i}$$v_\mathrm{drift}^2$ $\sim$ 22 eV \citep{tielens_2005}. This energy is not reached in the flow, even at the stagnation point where the drift velocity is maximal. Indeed, the impact energy $E$ should exceed the threshold energy with a factor of three to get effective sputtering of the grains. Therefore, we conclude that thermal sputtering of dust grains in the ionized flow is negligible.

\begin{figure}
\includegraphics[width=9cm]{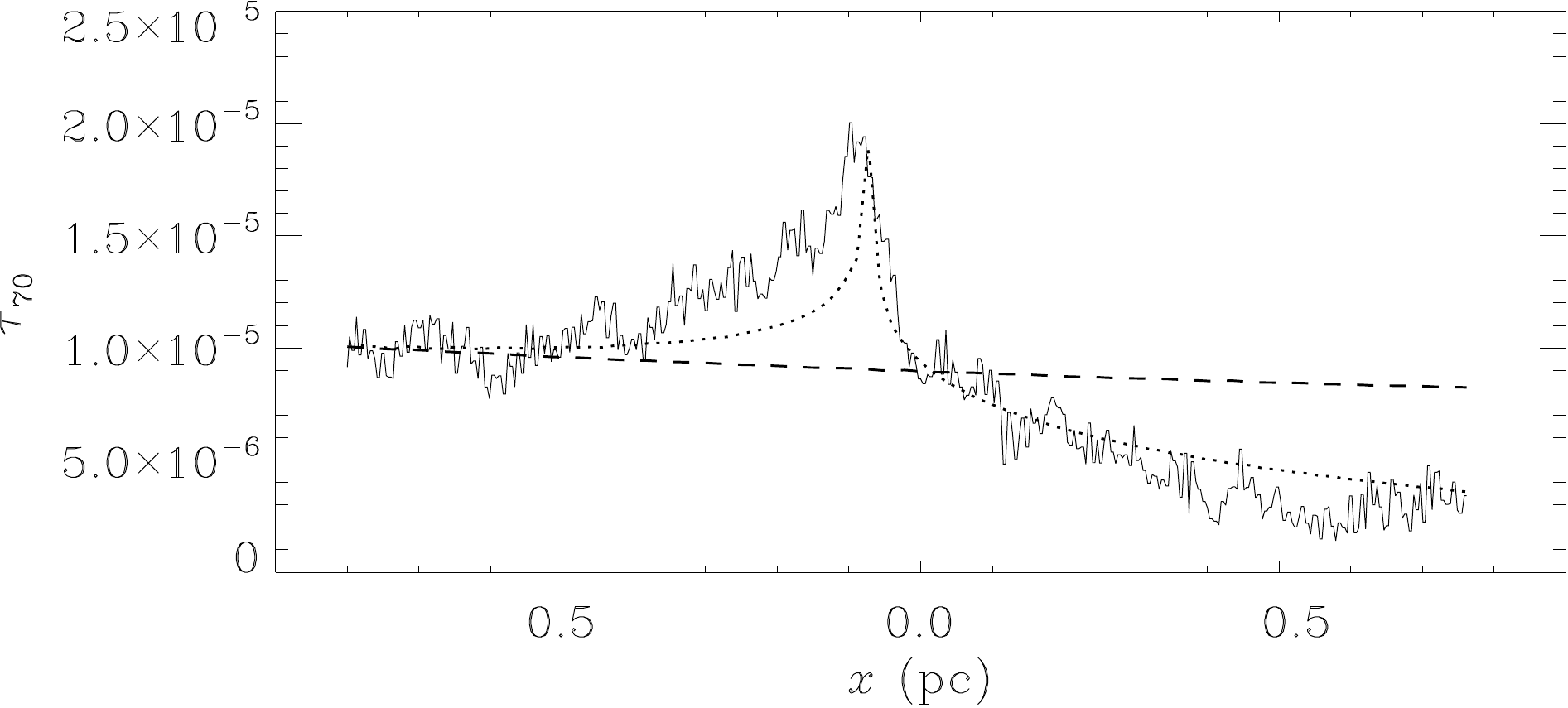}
\caption{A cross cut of the optical depth map in Fig. \ref{fig1:tempopt} taken horizontally through the stagnation point of the dust wave (solid). Overplotted is a model without radiation pressure force (dashed) and a model with radiation pressure force (dotted).}
\label{fig:crosscut}
\end{figure}

We calculate the velocity of the dust and gas at each $xyz$ point in space. We consider the problem to be axisymmetric along the $y$ and $z$ axes. Subsequently, we collapse the three-dimensional data cube along the line of sight and estimate the density increase using continuity: $\rho_\mathrm{d} v_\mathrm{d}$ = $C$, where $C$ is a constant. The result for grains with $\kappa_\mathrm{rp}$ = 6.7 $\times$ 10$^{3}$ cm$^2$ g$^{-1}$ is then compared against the observations in Fig. \ref{fig:crosscut}. We scale the model to the observed value of $\tau_\mathrm{70}$ at $x$ = 0.8 pc. In this way, the model gives an absolute increase in density along the line of sight. In Fig. \ref{fig:crosscut} we also show the solution of the model without radiation pressure. In this case, the dust will be accelerated as it is dragged by the gas and will follow the gas velocity law given by Eq. \ref{eq:velocity}. When the radiation pressure force is included (i.e., Eq. \ref{eq:eqmotiondust}), the dust will pile up in front of the star, resulting in a peak optical depth at $x$ = 0.1 pc. The observed optical depth at 0.5 $\textgreater$ $x$ $\textgreater$ 0.1 pc is significantly above the model value, which we attribute to the presence of a population of smaller grains which are deflected at larger radii. At $x$ $\textless$ 0 pc, the optical depth is lower compared to the model without radiation pressure force. This cavity can not be explained by mere acceleration of dust past the star (see Fig. \ref{fig:dust_gas_trajectories}). The decrement in optical depth is only reproduced by the model if we take trajectories with $\left| z \right|$ $\textless$ 0.5 pc into account. In other words, the observed optical depth at 70 $\mu$m (Fig. \ref{fig1:tempopt}) only traces dust at a distance of $r$ $\textless$ 0.5 pc from $\sigma$ Ori AB. This is because the 24 $\mu$m and 70 $\mu$m emission traces dust grains at a temperature of $\sim$ 30-100 K. 

In summary, we attribute the main peak in the optical depth map to large grains with $\kappa_\mathrm{rp}$ $\textgreater$ 1 $\times$ 10$^{4}$ cm$^2$ g$^{-1}$ ($a$ $\textgreater$ 1500 \AA\ for silicates). The enhanced optical depth in front of the main dust wave is then a result of grains with  $\kappa_\mathrm{rp}$ $\textless$ 1 $\times$ 10$^{4}$ cm$^2$ g$^{-1}$ and is tracing smaller dust grains or different grain composition (graphite versus silicate). Therefore, dust waves (as well as bow waves) are natural grain sorters separating dust grains according to their radiation pressure opacities.

\subsection{Coulomb interactions}\label{sec:charged_flow}

Thus far we have neglected grain charging of the dust grains. The charge of an interstellar grain is determined by photo-ionization and positive ion recombination balanced with negative electron recombination. Under typical ISM conditions, this results in a positive grain charge due to the harsh conditions of the interstellar radiation field combined with low electron densities. In an \HII region like IC 434, the electron density far exceeds that of the diffuse ISM. A good insight in the local physical parameters is needed in order to constrain the nature and magnitude of the grain charge.

The photo-ejection rate $J_\mathrm{pe}$ is dependent on the incident radiation field and the photo-ionization yield $Y_\mathrm{ion}$,

\begin{equation}
\label{eq:photorate}
J_\mathrm{pe} =  \sigma_\mathrm{d} \int_{\nu_{Z_\mathrm{d}}}^{\nu_\mathrm{max}} \frac{J (\nu)}{h\nu} Q_\mathrm{abs} Y_\mathrm{ion} (Z_\mathrm{d},\nu) d\nu,
\end{equation}

\noindent where $J$ represents the mean radiation intensity calculated using a O9.5V Kurucz model atmosphere \citep{kurucz_1993} and $Q_\mathrm{abs}$ is the absorption efficiency of the grain material. We take $Q_\mathrm{abs}$ for silicates from \cite{laor_draine_1993}. The photo-ionization yield $Y_\mathrm{ion}$ has been measured in laboratory experiments for energies up to 20 eV \citep{abbas_2006}, but experiments with higher energy photons are lacking and are very uncertain. \cite{Weingartner_2006} investigated the role of extreme ultraviolet radiation and X-rays and concluded that higher energy photons do not contribute significantly to the charge state of dust grains in \HII regions ionized by OB stars. We follow \cite{weingartner_2001} and take the ionization yield $Y_\mathrm{ion}$ as has been described there. 

The photo-ejection rate is balanced by the electron recombination rate. Assuming a Maxwellian velocity distribution of the gas, the collisional rate $J_\mathrm{e}$ between electrons and a grain of positive charge $Z_\mathrm{d}$ is estimated with \citep{tielens_2005}
 
\begin{equation}
\label{eq:recomrate}
J_\mathrm{e} (Z_\mathrm{d}) = n_\mathrm{e} s_\mathrm{e} \left(\frac{8kT}{\pi m_i}\right)^{1/2} \pi a^2 \left(1 + \frac{|\nu|}{\epsilon}\right)\left[1 + \left(\frac{2}{\epsilon + 2|\nu|}\right)^{1/2}\right].
\end{equation}

\noindent Here we define the reduced temperature, $\epsilon = akT/q_\mathrm{e}^2$ (with $q_\mathrm{e}$ being the charge of the electron), and the charge ratio between the dust and an electron, $\nu = Z_\mathrm{d}/q_\mathrm{e}$. If $\nu$ $\textless$ 0, the collisional rate increases because of electrostatic focussing. The charge distribution function is then calculated with Eq. \ref{eq:photorate} and \ref{eq:recomrate},

\begin{equation}
\label{eq:dragforce}
\frac{f(Z_\mathrm{d} + 1)}{f(Z_\mathrm{d})} = \frac{J_\mathrm{pe}(Z_\mathrm{d})}{J_\mathrm{e}(Z_\mathrm{d} + 1)},
\end{equation}

\noindent where $f$ is the fractional abundance of a dust grain with charge $Z_\mathrm{d}$. The maximum charge of a dust grain is calculated at a distance $d$ =  0.1 pc from the star. At these conditions, the peak of the charge distribution function lies at $Z_\mathrm{d}$ = 195 for a 300 \AA\ grain, whereas for a 3000 \AA\ grain this is $Z_\mathrm{d}$ = 1809.

\begin{figure*}
\centering
 \subfigure{\includegraphics[width=7.8cm]{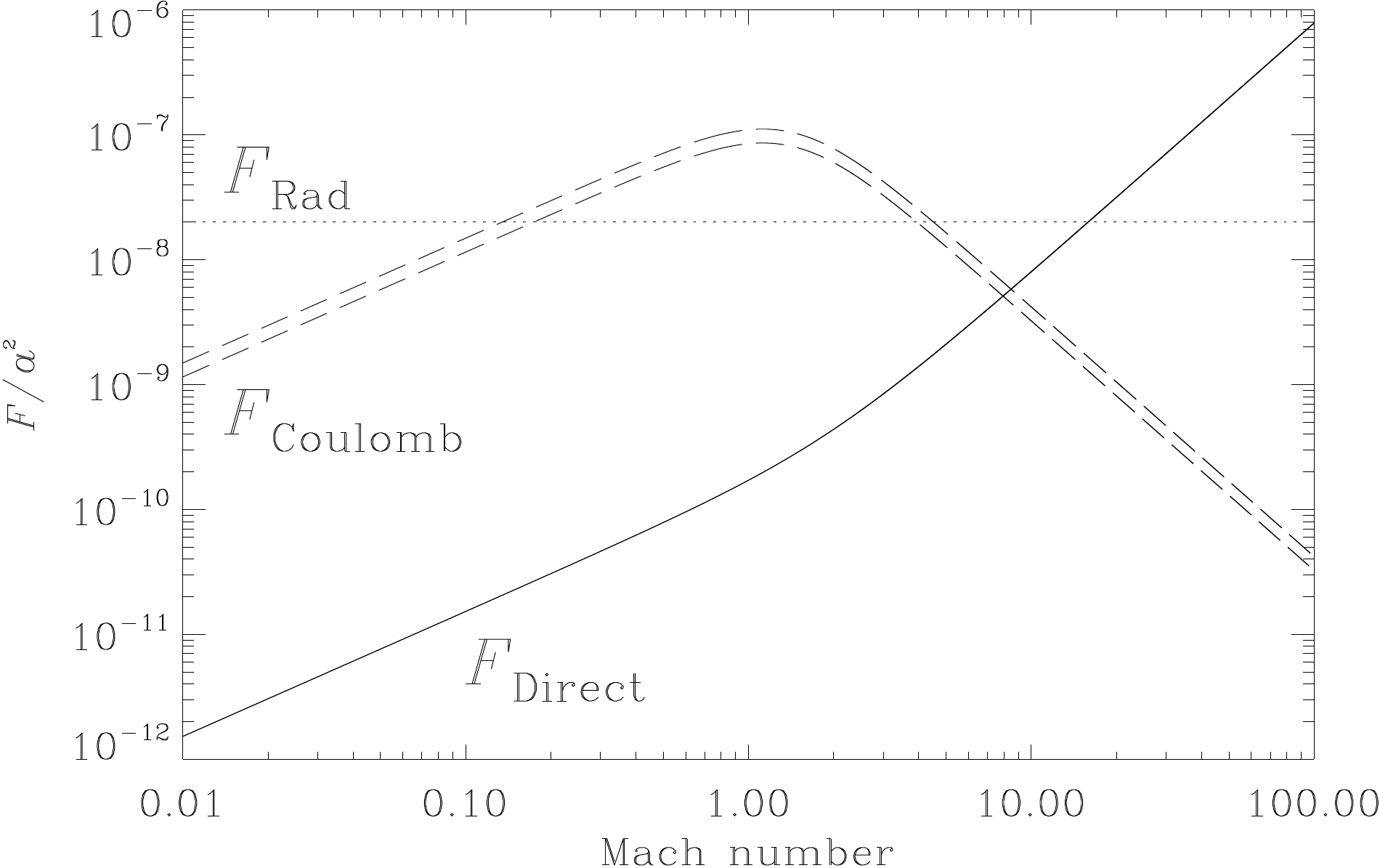}\label{fig:plasmavsdirect}}
 \qquad
 \subfigure{\includegraphics[width=8cm]{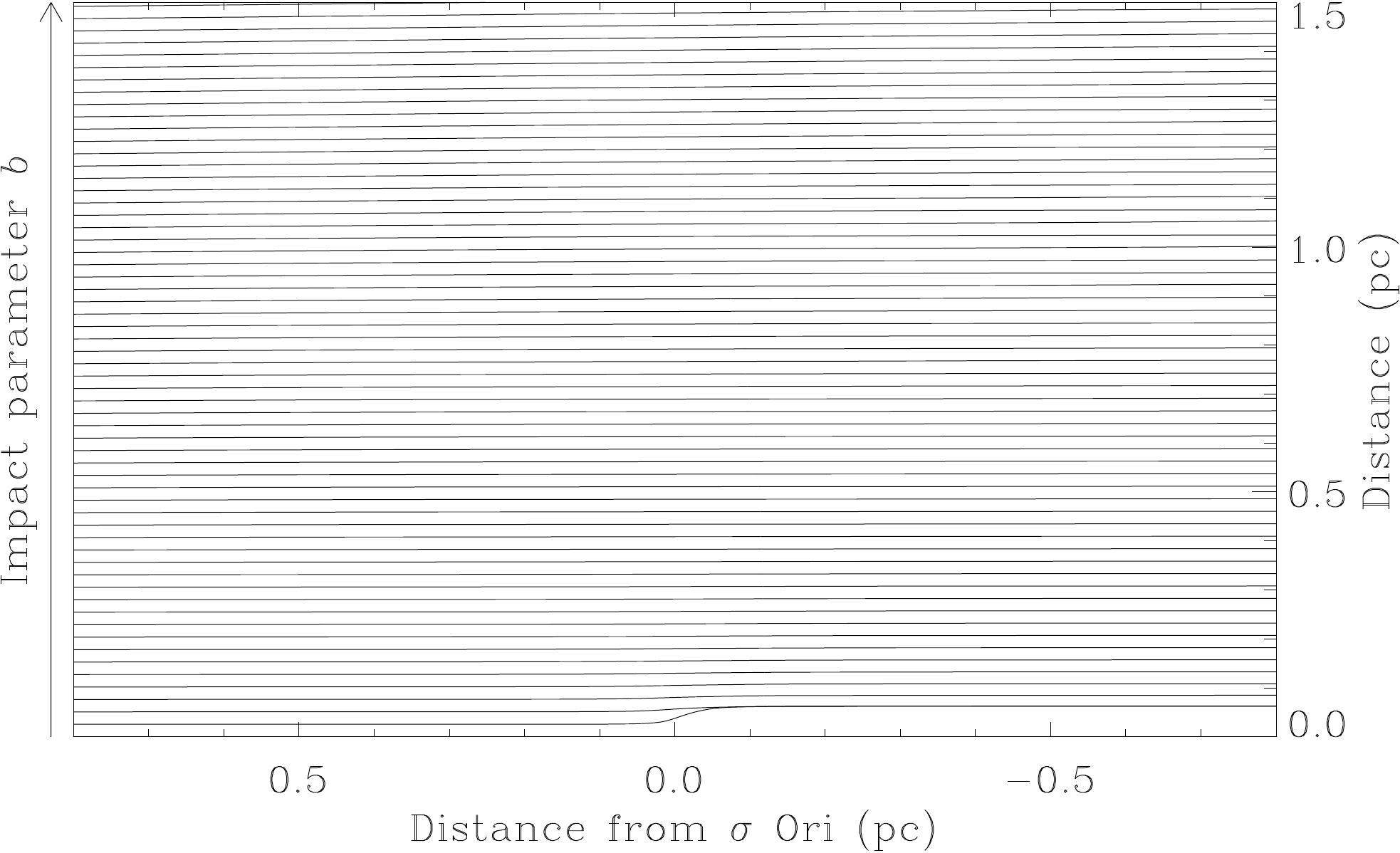}\label{fig:xy_plasma}}
\caption{The dependence of the direct drag force and Coulomb drag force on drift velocity and the structure of the photo-evaporation flow with the inclusion of Coulomb drag. {\em Left}: Direct drag (solid line) and the Coulomb drag on a 300 \AA\ dust grain with charge $Z_\mathrm{d}$ = 195 (upper dashed line) and a 3000 \AA\ dust grain with charge $Z_\mathrm{d}$ = 1809 (lower dashed line). Plotted is the dependence of the drag forces on the relative velocity between the gas and the dust, expressed in the Mach number $\mathcal{M}$ = $v_\mathrm{drift}$/$c_\mathrm{s}$. The dotted horizontal line represents the radiation pressure force at $r$ = 0.1 pc. {\em Right}: Calculated trajectories of a dust grain with $\kappa_\mathrm{rp}$ = 6.7 $\times$ 10$^{3}$ including the Coulomb drag force. The grains only decouple from the gas very close to the star (i.e., $d$ $\textless$ 0.025 pc).}
\label{fig:plasma_vs_direct}
\end{figure*}

A charged grain will not only have direct collisions with the gas (direct drag), but will also have long range Coulomb encounters with both electrons and ions (Coulomb drag). In addition, a grain will gyrate around the local magnetic field  through the Lorentz force if it has a velocity component perpendicular to the field lines. The magnitude of both the Coulomb and the direct drag force is highly dependent on $v_\mathrm{drift}$. At subsonic speeds, both direct and Coulomb drag forces are proportional to $v_\mathrm{drift}$. In the supersonic regime, the direct drag force increases with $v_\mathrm{drift}^2$, while the Coulomb drag is proportional to $v_\mathrm{drift}^{-2}$. Figure \ref{fig:plasma_vs_direct} relates the drag forces for a charged particle at 0.1 pc as a function of the Mach number $\mathcal{M}$ = $v_\mathrm{drift}$/$c_\mathrm{s}$. In Sec. \ref{sec:basic}, we have shown that the maximum drift speed in the flow is $\mathcal{M}$ $\sim$ 5. As a result, it seems not justified to neglect the Coulomb drag force as it should dominate the total drag force throughout the entire flow.

Figure \ref{fig:plasma_vs_direct} also shows the radiation pressure force at 0.1 pc. Particles with 0.1$\mathcal{M}$ $\textless$ $v_\mathrm{drift}$ $\textless$ 5 $\mathcal{M}$ will still be coupled to the gas at this distance. Figure \ref{fig:xy_plasma} shows trajectories of a dust grain with $\kappa_\mathrm{rp}$ = 6.7 $\times$ 10$^{3}$. Only at $d$ $\textless$ 2.5 $\times$ 10$^{-2}$ pc, does the radiation pressure force become larger than the drag force and the dust will decouple from the gas. In contrast, particles with $b$ $\textgreater$ 2.5 $\times$ 10$^{-2}$ pc stay coupled to the gas and will flow past $\sigma$ Ori AB without being deflected since the radiation pressure is not able to overcome the Coulomb drag force at these distances.

The observed shape of the dust wave in Fig. \ref{fig:dust_trajectories} can only be reproduced if $v_\mathrm{d}$ at the wave front parallel and perpendicular to the flow are of same order. With the inclusion of the Coulomb drag force, decoupling occurs at $d$ = 2.5 $\times$ 10$^{-2}$ pc for large grains which does not agree with the observations (decoupling at $d$ = 0.1 pc. In contrast, we have seen that by considering direct collisions only the observations are well reproduced. A further discussion on the exclusion of the Coulomb drag force is given in section \ref{sec:discussion}.

\subsection{Bow shock scenario}\label{sec:bowshock}
 
One may question in what way the dust wave surrounding $\sigma$ Ori AB differs from a bow shock (e.g., \citet{van_buren_1990,kaper_1997}). According to our definition stated in Sec. \ref{sec:structure}, a dust wave increases the number density of dust and will therefore solely emit at IR wavelengths with respect to the background emission, whereas a bow shock will shock both gas and dust and it is therefore possible to detect a bow shock in gas emission lines as well as IR wavelengths. This section elaborates on the gas emission expected from the structure if we were to treat is as a bow shock.

Equating the momentum flux of the stellar wind with the ram pressure of the star moving through the ISM, the stand-off distance $r_\mathrm{s}$ of a bow shock is given by \citep{van_buren_1988}

\begin{equation}
\label{eq:standoff}
r_s = 1.78 \times 10^3 \sqrt{\frac{\dot{M} v_\infty}{\mu_\mathrm{H} n_\mathrm{H} v_\mathrm{\star-\mathrm{ISM}}^2}} \textrm{                   pc},
\end{equation}

\noindent where $\dot{M}$ and $v_\infty$ are the mass-loss from the star in $M_{\odot}$ yr$^{-1}$ and the terminal velocity of the stellar wind in km s$^{-1}$; $\mu_\mathrm{H}$ and $n_\mathrm{H}$ are the mean mass per hydrogen nucleus (= 0.61 for a fully ionized medium) and the hydrogen gas density in cm$^{-3}$; and $v_\mathrm{\star-\mathrm{ISM}}$ is the relative velocity of the star with respect to the ISM in km s$^{-1}$. Plugging in numbers for $\sigma$ Ori AB \citep{najarro_2011} gives $r_\mathrm{s}$ = 8 $\times$ 10$^{-3}$ pc (using $v_\mathrm{\star-\mathrm{ISM}}$ = 50 km s$^{-1}$, $\mu_\mathrm{H}$ = 0.61, $\dot{M}$ = 2.0 $\times$ 10$^{-10}$ $M_\odot$ yr$^{-1}$ and the density profile for $n_\mathrm{H}$ derived in Sec. \ref{sec:emissionmeasure}). This radius does not coincide with the observed peak emission at 0.1 pc.

Within the bow shock scenario, the gas emission should coincide with the peak of the IR dust emission if we assume the grains to be coupled to the gas. We note that large grains could potentially cross the shocked ambient gas, traversing into the shocked wind region due to their high momentum (Cox et al. 2013, in preparation). This would displace the peak IR emission from the peak gas emission if these grains dominate the size distribution. Here, we assume efficient gas-dust coupling in the shocked region. The observed stand-off distance is only reproduced by increasing the wind momentum flux $\dot{M}v_\infty$ by a significant amount (a factor of $\sim$ 300). For now, we adopt the wind parameters from \citet{howarth_1989} in order to estimate the gas density and hence the expected emission measure (EM) coming from a bow shock. 

The temperature increase in a fully ionized medium is given by \citep{tielens_2005}

\begin{equation}
\label{eq:postshock_t}
T_1 = 1.4\times10^1 \left(\frac{v_\mathrm{\star-\mathrm{ISM}} }{\textrm{ km s}^{-1}}\right)^{2} \mathrm{           K},
\end{equation}

\noindent where $T_\mathrm{1}$ represents the postshock temperature. At $T_\mathrm{1}$ $\simeq$ 10$^4$ K, the gas is cooled by permitted and (semi-)forbidden transitions, while collisional de-excitation is unimportant. In these circumstances, the post shock column density has to exceed the cooling-column density $N_\mathrm{cool}$ for a radiative shock to develop:

\begin{equation}
\label{eq:coolingcolumn}
N_\mathrm{cool} \simeq 8.2\times10^{8} \left(\frac{v_\mathrm{\star-\mathrm{ISM}} }{\textrm{ km s}^{-1}}\right)^{4.2} \mathrm{           cm^{-2}}.
\end{equation}

\noindent Given the expected relative velocity $v_\mathrm{\star-\mathrm{ISM}}$ = 50 km s$^{-1}$ and $T_\mathrm{1}$ = 3.5 $\times$ 10$^4$ K, $N_\mathrm{cool}$ = 1.1 $\times$ 10$^{16}$ cm$^{-2}$. The shape of a bow shock near the stagnation point can be approximated by $x = y^2$/$3l$ \citep{van_buren_1990}, where $x$ and $y$ are the coordinates parallel and perpendicular to the direction of motion. The column density of the swept-up material at the stand-off position can then be estimated using 

\begin{equation}
\label{eq:columndensity}
N_\mathrm{H} = 6.18 \times 10^{21} \sqrt{\frac{\dot{M} n_\mathrm{H} v_\infty}{\mu_\mathrm{H}  v_\star^2}} \mathrm{                 cm^{-2}}.
\end{equation}

\begin{figure}         
\centering
\includegraphics[width=9cm]{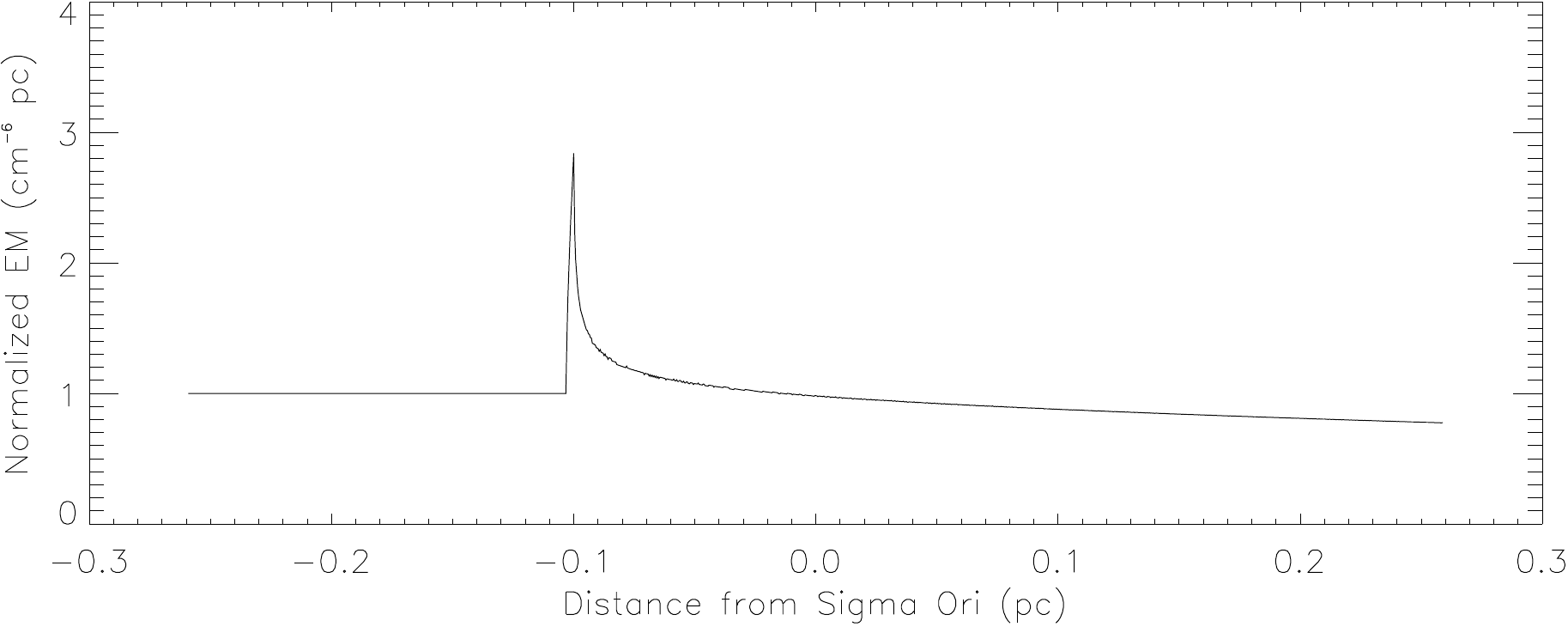}
 \caption{The emission measure from the gas expected from a bow shock as described in \citet{wilkin_1996}, normalized to the expected background emission measure of the entire flow. The computed emission profile at resolution 0.26" (the resolution of the KPNO H$\alpha$ image) has been convolved with a Gaussian of width 1" to account for (typical) astronomical seeing at the wavelength of H$\alpha$. A cut through the stagnation point and the star is shown, adopting wind parameters from \citet{howarth_1989} and ISM parameters calculated in section \ref{sec:flow}.}
 \label{fig:em}
\end{figure}

\noindent Here, the same units are used as in Eq. \ref{eq:standoff}. Using the wind parameters of \citet{howarth_1989}, the post-shock column density is estimated to be 4.6 $\times$ 10$^{18}$ cm$^{-2}$ and therefore we consider the shock to be radiative. A simple analytic expression for the thickness $\delta$ in pc of a radiative shock at the stagnation point is given through mass and momentum conservation \citep{van_buren_1990},

\begin{equation}
\label{eq:column}
\delta = 16.6\cdot \gamma_\mathrm{I} T_\mathrm{I} \sqrt{\frac{\dot{M} v_\infty}{\mu_\mathrm{H}^{3} n_\mathrm{H} v_\star^6}}  \textrm{         pc},
\end{equation}

\noindent where $\gamma_\mathrm{I}$ is the ratio of specific heats in the preshock gas (= 5/3) and $T_\mathrm{I}$ its temperature (= 7500 K). We calculate a thickness $\delta$ = 0.01 pc; this corresponds to an angular size of 6$\arcsec$ at a distance of 334 pc and a scale size along the line of sight of 0.1 pc, which indicates that the shock is resolved at the resolution of the KPNO H$\alpha$ image (0.26"). The shape and emission of a bow shock can be estimated as described in \citet{wilkin_1996} assuming that the post-shock cooling is efficient (thin-limit approximation). The result is plotted in Fig. \ref{fig:em}, where we have calculated the EM at the resolution of the KPNO H$\alpha$ image. We can predict the increase in emission measure seen at the wavelength of H$\alpha$ by convolving the computed profile with a Gaussian kernel of 1" in width (a typical value for astronomical seeing in the optical). At these conditions, we expect an increase of nearly 300\% at the stagnation point with respect to the background emission measure of the entire photo-evaporating flow at optical wavelengths. 

The Spitzer/IRS spectra provides insight into the gas structure of the potential shock. Prominent lines in the 15-35 $\mu$m wavelength range are the S[III] 33.4 $\mu$m and S[III] 18.7 $\mu$m infrared fine-structure lines. Given the high critical densities and low excitation energy (compared to the electron temperature) of these lines, IR fine structure lines are good tracers of density variations in the emitting gas. Here, we have elected to use the 18.7 $\mu$m line because of its higher critical density (1.2 $\times$ 10$^{4}$ cm$^{-3}$) compared to the 33.4 $\mu$m line to exclude collisional de-excitation. As the gas density in the IC 434 photo-evaporative flow is well below the critical density, we expect a similar increase in the S[III] emission measure as plotted in Fig. \ref{fig:em}, albeit somewhat lower due to the lower resolution of the IRS spectrograph compared to the resolution used in computing the profile in Fig. \ref{fig:em} (1"). At 19 $\mu$m, the FWHM of the PSF of the IRS is $\sim$5" (from the IRS instrument handbook 5.0); this would shift the peak value of the expected emission measure down to 230\% the background value. However, in reality bow shocks are not so thin \citep[e.g.,][]{comeron_1998}, therefore the increase in emission measure may not be as pronounced as depicted in Fig. \ref{fig:em}, which should be treated as a limiting case. 

\begin{figure}
  \centering
 \subfigure{\label{fig:shassaregion}\includegraphics[width=9cm]{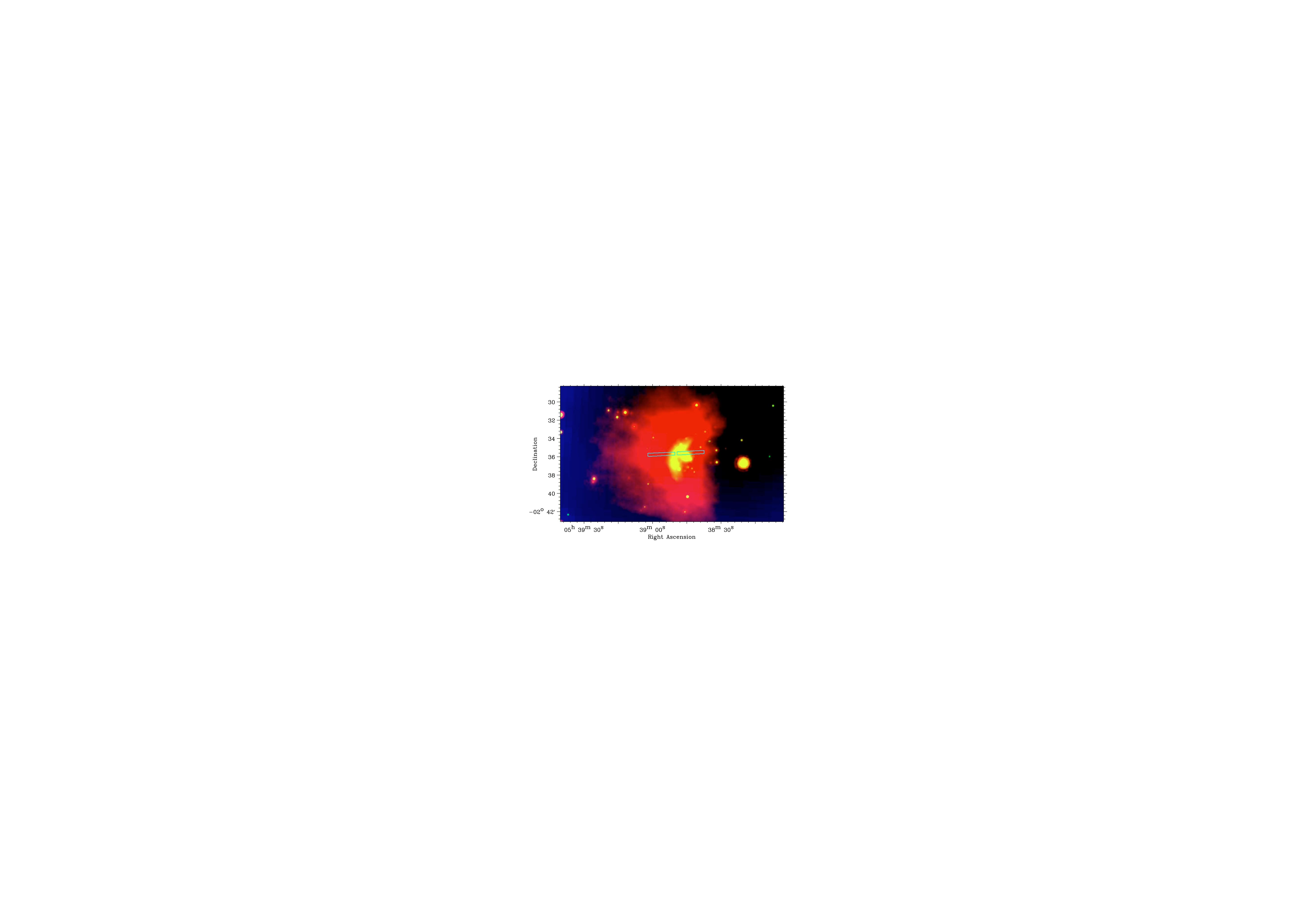}}
 \subfigure{\label{fig:spitzerline}\includegraphics[width=9cm]{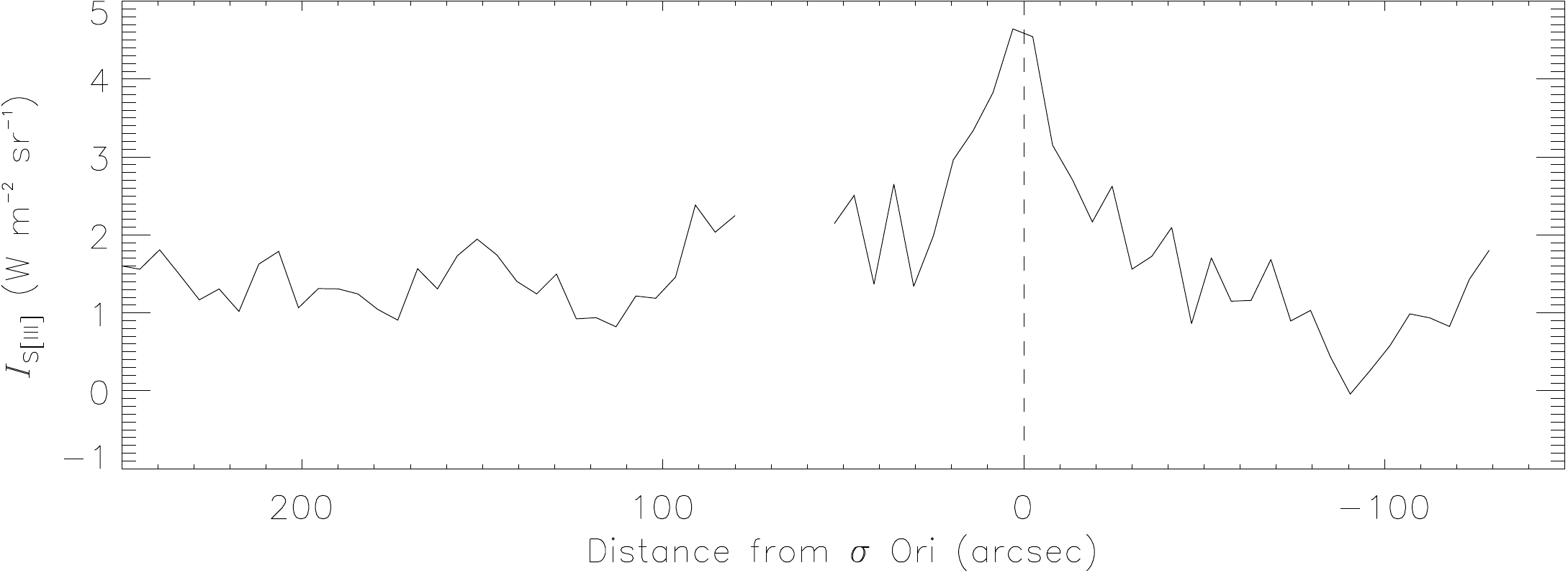}}
  \caption{{\em Upper}: 0.45${\degr}$ $\times$ 0.2${\degr}$ three color image of the IR emission around $\sigma$ Ori AB. Red is WISE-4 24 $\mu$m, green is WISE-3 12 $\mu$m and blue is H$\alpha$ from the SHASSA mission, smoothed over 5 pixels in order to remove star residuals. Overplotted are the Spitzer/IRS LL2 regions. {\em Lower}: Continuum subtracted and integrated intensity of the S[III] 18.7 $\mu$m line inside the IRS regions as a function of distance towards $\sigma$ Ori AB (a positive distance corresponds to the region in front of the star). The dashed line marks the position of $\sigma$ Ori AB. The IR arc is located at $\sim$ 0.1-0.4 pc (the peak IR value being at $\sim$ 0.1 pc), corresponding to 50-200 arcseconds in front of $\sigma$ Ori AB.}
 \label{fig1:spitzerregions}
\end{figure}

The covered regions of the IRS observations are overplotted in Fig. \ref{fig1:spitzerregions}. In the lower panel of Fig. \ref{fig1:spitzerregions}, the intensity of the S[III] 19 $\mu$m line is plotted as a function of distance from $\sigma$ Ori AB. We only detect gas emission close to $\sigma$ Ori AB, possibly originating from the immediate surroundings of the system. We do not detect a significant increase (i.e., \textgreater\ 3$\sigma$) in line emission at the position of the dust structure as is seen in the IR images (0.1 - 0.4 pc, or 50 - 200"). However, the IRS observations are noisy; high resolution imaging of the $\sigma$ Ori AB region is needed to conclude on the (non)-existence of a gaseous structure surrounding $\sigma$ Ori AB. We emphasize that the weak stellar wind as derived by \citet{najarro_2011} is not able to create a bow shock with an associated IR arc at the observed scale size. $\sigma$ Ori AB needs to have a powerful stellar wind such as measured by \citet{howarth_1989} to create an arc structure at the observed distance. This in turn should give strong increase in H$\alpha$ emission measure (or another gas tracer) which is yet to be observed. Here, we conclude that, inside the weak-wind scenario, the formation of a dust wave provides a very plausible alternative to the previously invoked bow shock designation.

\section{Discussion}\label{sec:discussion}

Within the dust wave scenario, the dust grain motions are affected by radiation pressure and drag force through interaction with gas particles. Our results indicate that the momentum transfer between the gas and dust is efficient at the start of the photo-evaporation flow. Small grains  ($a$ $\textless$ 1000 \AA) will initially be coupled to the gas, whereas larger grains ($a$ $\textgreater$ 1000 \AA) will not be accelerated to the gas velocity. As the combined dust-gas flow travels into the \HII region, the gas density decreases while the radiation pressure increases, which ultimately leads to a decoupling of gas and dust. The point of decoupling depends on the gas density and the dust grain properties, in particular the radiation pressure cross section per unit mass, $\kappa_\mathrm{rp}$. Close to $\sigma$ Ori AB, the number density of the dust particles have dropped to a level such that the total momentum transferred back to the gas is negligible. Therefore, the gas flows through the dust wave unhindered. Dust waves and bow waves can be distinguished by calculating if the dust is able to stop the gas along with it. This enters in our theory through Eq. \ref{eq:eqmotiongas}, where in the case of a dust wave the velocity of the gas ($v_\mathrm{g}$) is equal to the relative velocity between star and ISM ($v_\mathrm{\star-\mathrm{ISM}}$).

In our analysis, we have implicitly assumed that the space velocity of $\sigma$ Ori AB (15 km s$^{-1}$) represents the velocity between the L1630 molecular cloud and $\sigma$ Ori AB (i.e., the cloud is static with respect to the ISM). It is uncertain whether this is true: the Galactic rotation model used to calculate space velocities is an estimate and large discrepancies can occur. The other extreme (i.e., the cloud is static with respect to $\sigma$ Ori AB) will lower the velocity of the flow by a fixed offset of 15 km s$^{-1}$. This effect will only qualitatively influence our results in the form of a different size distribution of dust, as larger grains are needed to reach the projected distance of the dust wave (0.1 $\textgreater$ $d$ $\textgreater$ 0.4 pc). Similarly, as we have noted in Sec. \ref{sec:intro}, the distance towards $\sigma$ Ori AB is uncertain. This would only change the projected distance between the star and cloud and not affect our main conclusions.

One may question the uniqueness of the dust wave around $\sigma$ Ori AB. In order to create a dust wave (or a bow wave), the point where dust grains are stopped due to radiation pressure should exceed the stand-off distance of a bow shock, i.e.,  $r_\mathrm{min}$ $\textgreater$ $r_\mathrm{s}$. In this way, radiation pressure will act on the dust instead of being shocked by the stellar wind. It is therefore critical to have a good understanding of the mechanism driving stellar winds in order to constrain wind parameters from host stars. 

\begin{figure}
\includegraphics[width=9cm]{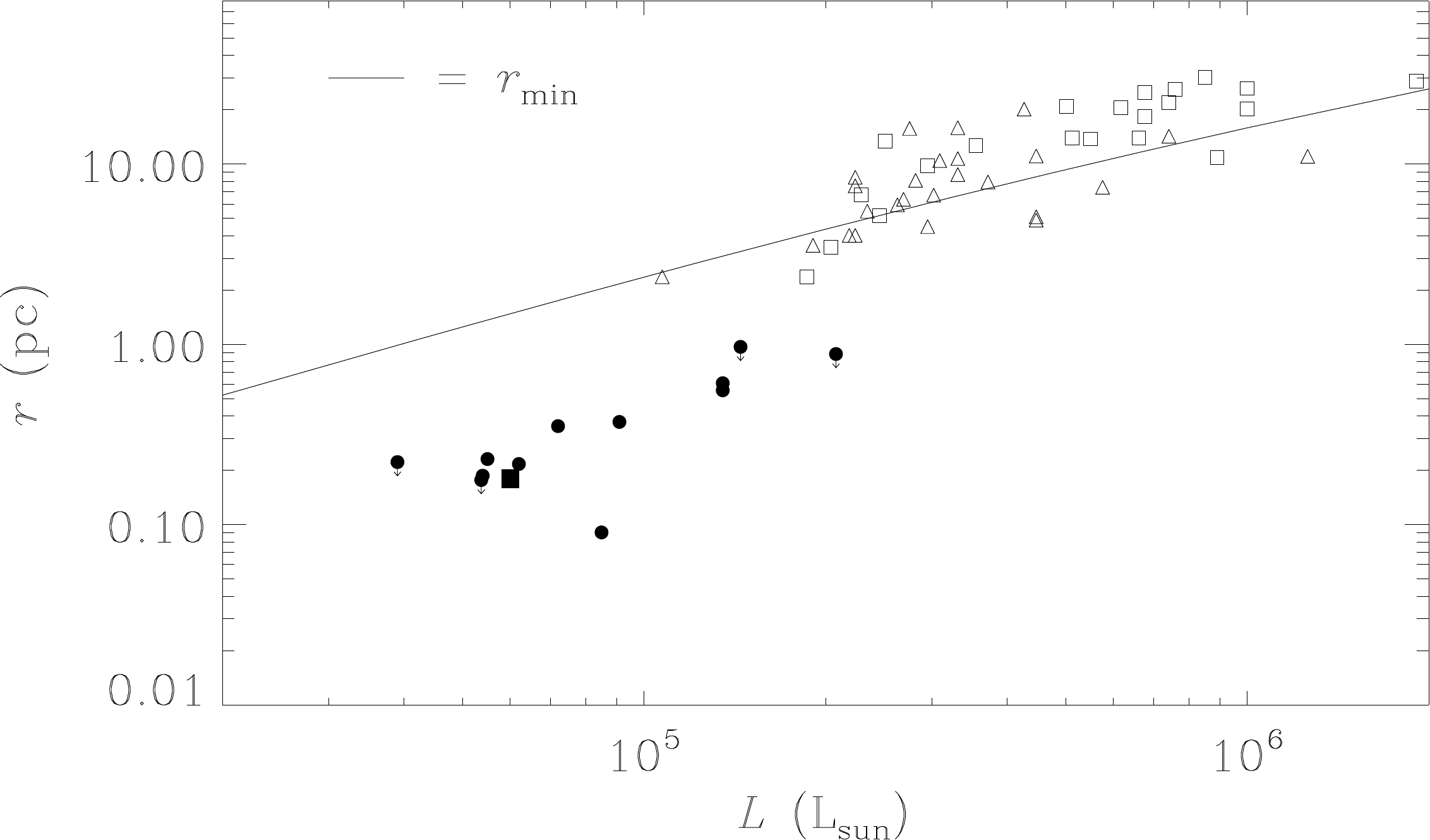}
\caption{The solid curve shows $r_\mathrm{min}$ for dust grains with $\kappa_\mathrm{rp}$ = 6.7 $\times$ 10$^{3}$ cm$^2$ g$^{-1}$ plotted as a function of luminosity $L$, assuming typical parameters for the stellar velocity (= 10 km$^{-1}$) and an ISM density similar to the WNM (= 0.5 cm$^{-3}$). Overplotted are Galactic O-stars (open squares) and B-supergiants (open triangles) where wind parameters are taken from \citet{puls_1996} and \citet{crowther_2006}. The filled circles are weak-wind objects from \citet{bouret_2003}, \citet{martins_2004} and \citet{najarro_2011}. $\sigma$ Ori AB is marked as the filled square symbol.}
\label{fig:lum_rmin}
\end{figure}

We evaluate the situation where a star is moving trough the Warm Neutral Medium (WNM; $n_\mathrm{H}$ $\approx$ 0.5 cm$^{-3}$) with a speed of 10 km s$^{-1}$ and compare numbers for $r_\mathrm{min}$ and $r_\mathrm{s}$. Figure \ref{fig:lum_rmin} shows the relation $r_\mathrm{min}$ versus luminosity $L$. Overplotted are predictions for the stand-off distance $r_\mathrm{s}$ using observed wind parameters (Eq. \ref{eq:standoff}). Galactic O-stars with spectral type earlier than O6 and early type B-supergaints show $r_\mathrm{min}$ $\textless$ $r_\mathrm{s}$. In general, these stars have a too powerful wind to create a dust wave. In contrast, the weak winds observed for dwarf stars with spectral type later than $\sim$ O6-O7 are able to create a dust wave ($r_\mathrm{min}$ $\textgreater$ $r_\mathrm{s}$). 

\begin{figure}
\includegraphics[width=9cm]{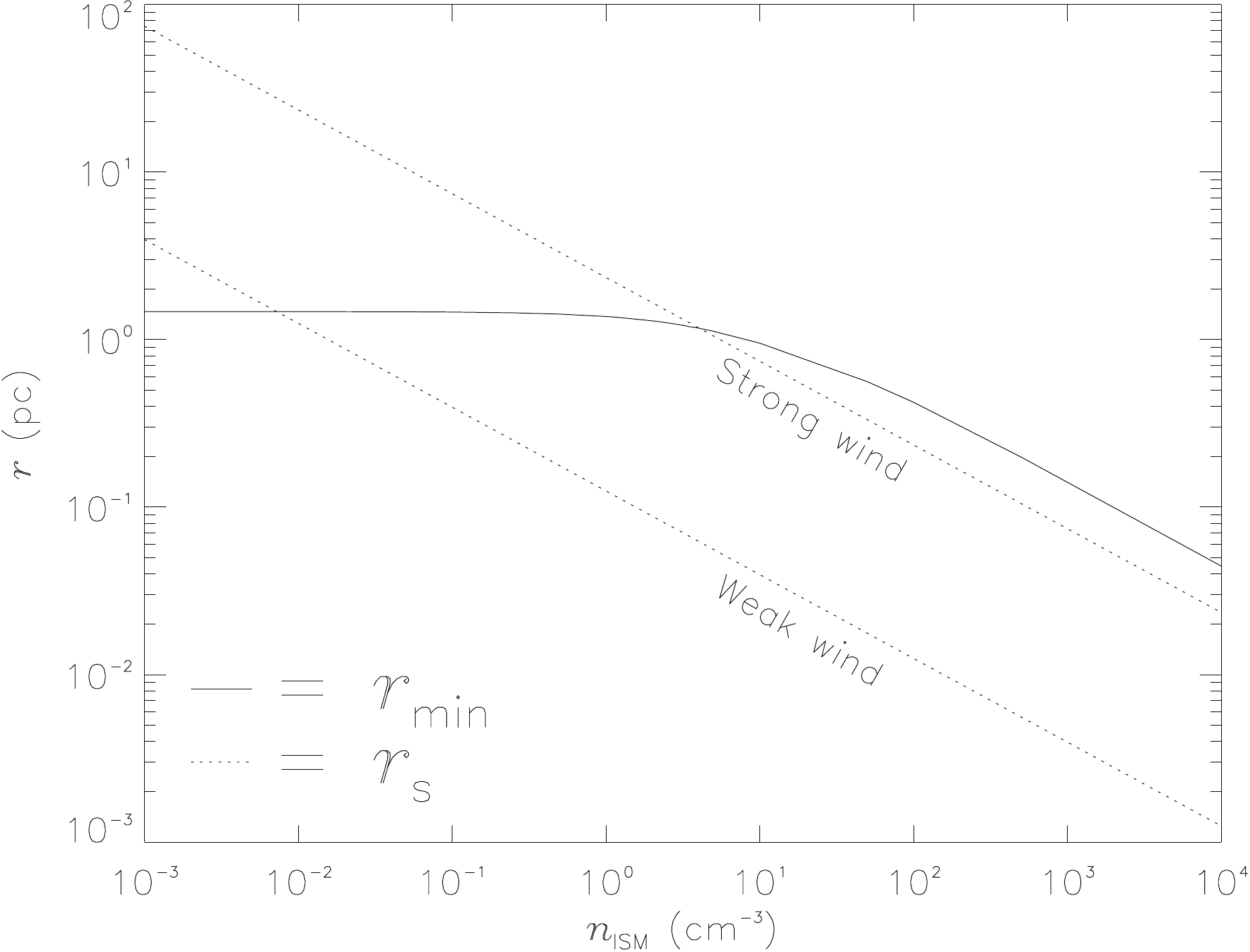}
\caption{$r_\mathrm{min}$ and $r_\mathrm{s}$ as a function of ISM number density $n_\mathrm{ISM}$. The stand-off distance $r_\mathrm{s}$ is plotted using wind parameters from \citet{najarro_2011} (lower dotted line) and \citet{vink_2000} (upper dotted line). $r_\mathrm{min}$ reaches a maximum at $n_\mathrm{ISM}$ $\textless$ 10 cm$^{-3}$ where the drag force $F_\mathrm{d}$ becomes negligible. $r_\mathrm{s}$ and $r_\mathrm{min}$ show the same dependency on ISM density for $n_\mathrm{ISM}$ $\textgreater$ 10 cm$^{-3}$.}
\label{fig:lum_poly}
\end{figure}

We can generalize these conclusions by investigating the relation of $r_\mathrm{min}$ and $r_\mathrm{s}$ with ISM number density ($n_\mathrm{ISM}$) and stellar velocity ($v_\mathrm{\star}$). At low densities, the drag force $F_\mathrm{d}$ in Eq. \ref{eq:eqmotiondust} becomes negligible and $r_\mathrm{min}$ will be insensitive to a further decrease in density of the ambient medium. This is reflected in Fig. \ref{fig:lum_poly} at the point where the curve for $r_\mathrm{min}$ turns over. At high densities, the dependency between $r_\mathrm{min}$ and $n_\mathrm{ISM}$ is not straightforward and we opted to solve this numerically. The curves for $r_\mathrm{min}$ and r$_\mathrm{s}$ run parallel at high densities ($r_\mathrm{min}$ $\propto$ $r_\mathrm{s}$ $\propto$ $n_\mathrm{ISM}^{-0.5}$). The same relation between $r_\mathrm{min}$ and $n_\mathrm{ISM}$ is seen at different stellar velocities but for a subtle difference: at low $v_\mathrm{\star}$, the point where the curve for $r_\mathrm{min}$ turns over will shift to lower densities as the drag force can not be neglected at small velocities. The inverse holds for high $v_\mathrm{\star}$. The stand-off distance $r_\mathrm{s}$ is plotted in Fig. \ref{fig:lum_poly} using wind parameters from \citet{najarro_2011} (weak-wind scenario) and \citet{vink_2000}, respectively. It is clear that it is difficult to observe a dust wave around a star with a powerful wind described by \citet{vink_2000}. This is not the case for weak-wind stars. In particular, Fig. \ref{fig:lum_poly} reveals that creating a dust wave is most efficient around relatively slow moving stars where the drag between gas and dust in the ISM is still important. For a typical stellar velocity of $v_\star$ = 10 km s$^{-1}$ this criterium holds for $n_\mathrm{ISM}$ $\textgreater$ 10 cm$^{-3}$.  For a slowly moving star ($v_\star$ = 1 km s$^{-1}$) this drops to  $n_\mathrm{ISM}$ $\textgreater$ 0.1 cm$^{-3}$. In summary, it is possible to create a dust wave around virtually all weak-wind stars. Only high-velocity runaway stars moving through low density ISM regions will more likely create a bow shock rather than a dust wave. 
 
The Galactic sample of weak-wind candidates contains 22 sources, from which 5 have detected arc structures around them best observed at mid-IR wavelengths ($\sim$20 $\mu$m) \citep{gvaramadze_2012}: HD34078, HD48099, HD48279, HD149757 and HD216898. These structures are traditionally explained as bow shocks \citep{gvaramadze_2012,peri_2012}. We have investigated if these structures could be explained as a dust wave. In Appendix \ref{sec:appendix}, we compare the observed stand-off distance for these structures with (1) the expected stand-off distance within the bow shock scenario (both weak-wind and normal-wind) and (2) the location of a dust wave within the dust wave/bow wave scenario, as presented in this work. We conclude that the observed arc structures can only be explained by bow shocks if the stars have strong winds, close to the value predicted by the \citet{vink_2000} mass loss recipe (HD149757 may be an exception to this rule; see \citet{gvaramadze_2012}). Unfortunately, the mass loss rate of late type O-dwarf stars is still being debated, and until now no observations exist that could distinguish the above mentioned structures as stellar wind driven or radiation pressure driven; high-resolution imaging of gas tracers might help to resolve this issue. Here we emphasize that, within the weak-wind scenario, these arc structures are well described by a dust wave.

Each of the first four previously named weak-wind candidates with IR arcs are runaway-OB stars, however with moderate proper motions (25-40 km s$^{-1}$) except for HD34078 (150 km s$^{-1}$). Nevertheless, this seems to contradict the above discussion where we concluded that dust waves and bow waves are created most efficiently around slowly moving stars. This is obviously an observational bias: runaways provide a unique possibility to study massive stars individually, as they moved away from their often obscured formation sites. The lifetime of massive stars is in general too short to create a dust wave around stars with low space velocity. In contrast, lower mass stars intrinsically have a lower space velocity, a longer lifetime and can be studied individually. These stars would therefore be ideal candidates to observe a dust wave, but a detection will be hampered by limiting spatial resolution with current IR instrumentation.  

\begin{figure}
\includegraphics[width=9cm]{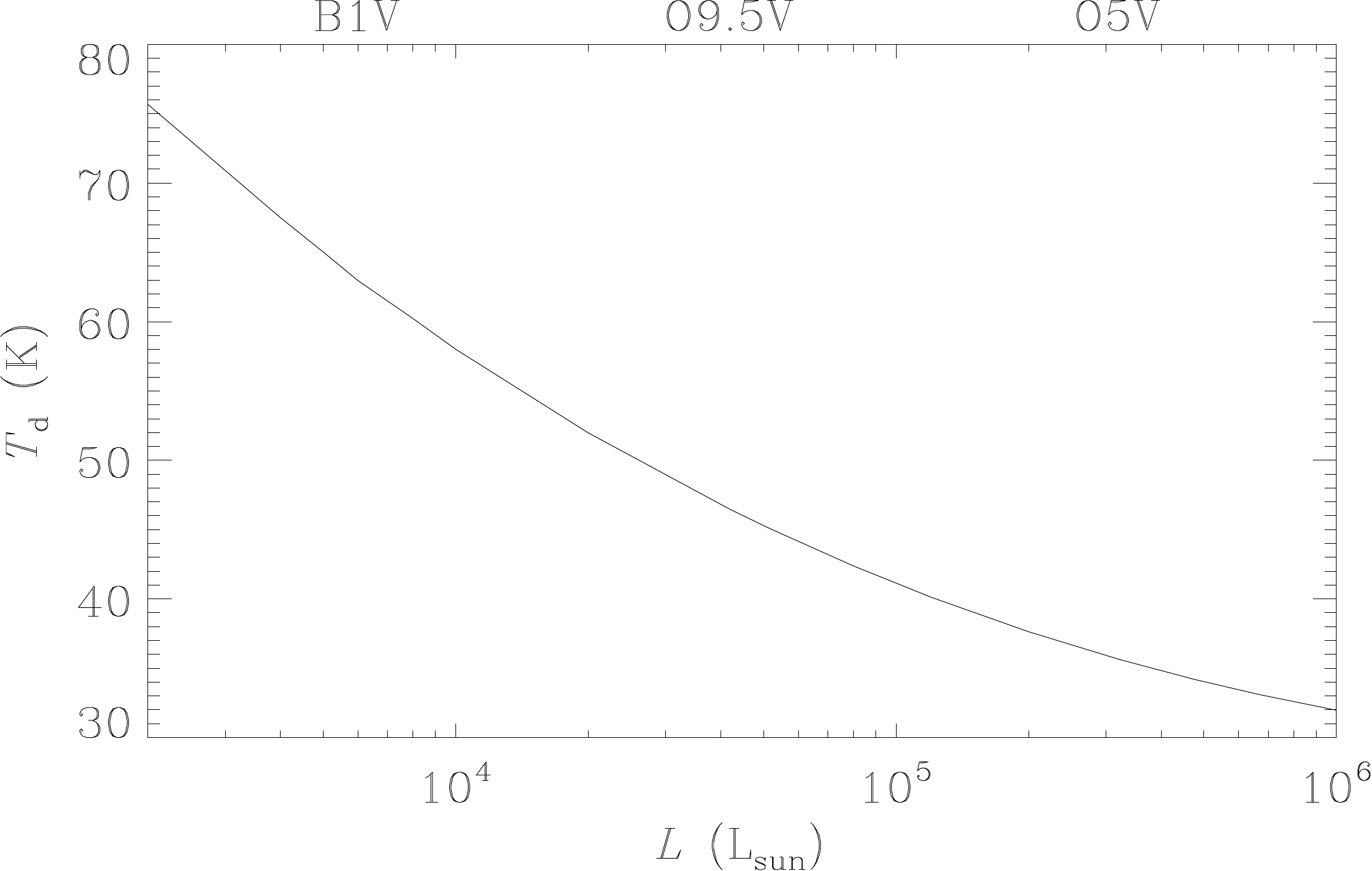}
\caption{Dust temperature $T_\mathrm{d}$ for silicate grains ($\kappa_\mathrm{rp}$ = 6.7 $\times$ 10$^{3}$ cm$^{2}$ g$^{-1}$), located at $r_\mathrm{min}$, plotted as a function of luminosity $L$. For comparison, several spectral types for main sequence stars are plotted on the top axis.}
\label{fig:lum_td}
\end{figure}

One of the key parameters to be able to detect a dust wave is the dust temperature. The temperature of silicates depends on the incident radiation field: $T_\mathrm{d}$ $\propto$ ($L/r_\mathrm{min}^2$)$^{1/6}$. Figure \ref{fig:lum_rmin} shows that $r_\mathrm{min}$ is roughly proportional to $L$, which leads to $T_\mathrm{d}$ $\propto$ (1/$L$)$^{1/6}$. Figure \ref{fig:lum_td} shows $T_\mathrm{d}$ as a function of $L$. In the WNM, dust waves around massive O-stars should be observed at FIR wavelengths, whereas dust waves around B-stars are best observed at mid-IR wavelengths. Clearly, these estimates depend on the choice of local physical parameters. The dust wave around $\sigma$ Ori AB is a striking example: due to the high density and the high relative velocity caused by the ionized flow, $r_\mathrm{min}$ decreases as compared to a similar structure in the WNM. Therefore, $T_\mathrm{d}$ rises to 70 K and the dust wave lights up at wavelengths detectable to WISE and Spitzer.

\subsection{Grain charge and Coulomb interaction}

The Coulomb interaction potential is originally derived by \citet{chandrasekhar_1943} for dynamical friction in a cluster of stars but redefined for a test particle confined in a plasma by \citet{spitzer_1962} and \citet{draine_salpeter_1979}. This theory treats each dust particle as being independent of one another, which is justified when the Debye screening length $\lambda_\mathrm{D}$ = ($kT/4\pi n_\mathrm{e})^{1/2}$ is smaller than the average intergrain distance $n_\mathrm{d}^{1/3}$. The grains inside the \HII region are expected to contain high positive charges. We have argued that the expected charge leads to a large coupling between the gas and dust through Coulomb interactions, which can not be reproduced with our model when compared with observations. Therefore, we have concluded that the Coulomb drag force in the photo-evaporation flow approaching $\sigma$ Ori AB can not be as efficient as is described in theory. We note that ionization yields for interstellar grains could be lower than expected. In Sec. \ref{sec:charged_flow}, we already discussed the uncertainty of photo-electric yields for higher energies, but even for lower energies, these yields remain uncertain. This can partly be attributed to the fact that these measurements have been made on flat surfaces of bulk material of astrophysical composition. Bulk yields can differ significantly from yields of individual dust grains \citep{draine_1978}. For example, yield curves for carbon and silicate as measured by \citet{willis_1973} and \citet{abbas_2006} differ by an order of magnitude. In addition, recent laboratory studies have revealed that high electric fields can spontaneously occur in solid films around cosmic dust analogs \citep{plekan_2011}. This is due to dipole alignment in the solid films and can create electrical fields of up to 10$^{10}$ V cm$^{-1}$. This can have a significant impact on the grain charge, depending on the orientation of the electrical field. 

It is also possible that we have underestimated the relative velocity between the star and material in the flow. A larger velocity would cause the dust to approach the star more closely and the Coulomb drag force will decrease by an order of magnitude. Specifically, if we increase the flow velocity by a factor 2 (i.e., $v_\mathrm{g}$ = 100 km s$^{-1}$), we need to increase the radiation pressure efficiency by a factor of 2, i.e., $\bar{Q}_\mathrm{rp}$ = 2, which is still comparable with the value from \citet{laor_draine_1993}. 

Aspects of grain charge have never been studied directly in the ISM and theory relies on idealized models. Laboratory studies show that there are still key uncertainties in both photo-electric yields of cosmic dust analogs as well as on electric fields of solid films. Furthermore, dust properties in \HII regions could be different from those in the diffuse ISM \citep{salgado_2012}. Dust waves will provide a unique environment to study the interaction of dust in an ionized environment.

\section{Conclusion}\label{sec:conclusions}

In this paper we have argued that the arc-shaped emission surrounding $\sigma$ Ori AB is a dust wave. Radiation pressure has created a structure ahead of the star where dust is being piled up, sorting the dust grains according to their radiation pressure opacities. A dust wave discriminates itself from a bow wave through the decoupling of the gas. We emphasize that to form the dust wave around $\sigma$ Ori AB, Coulomb interactions have to be unimportant to reach sufficient decoupling between dust and gas. This suggests that we may not fully understand the processes controlling dust charging in hot, ionized regions in space such as \HII regions.

We have shown that a dust wave, where radiation pressure acts on the surrounding medium, differs fundamentally from a classical bow shock, where the ram pressure of the star and ISM balance. It has already been proposed by \citet{van_buren_1988} that radiation-pressure-driven structures should exist around stars where the expected momentum flux from the stellar wind is low. No shock heating of gas is needed in the dust wave scenario, which explains the absence of gas emission lines in observations of bow shocks, in our case the S[III] observations. We note that observing bow shocks in the mid-IR is in general more efficient than in the optical. For example, typical conditions for a bow shock created by a massive runaway star moving through the ISM ($v_\star$ = 50 km/s, $n_\mathrm{H}$ = 1 cm$^{-3}$) give EM $\sim$ 25 cm$^{-6}$ pc \citep{van_buren_1993}, which is hardly detectable in present low-resolution all sky optical surveys. On the other hand, even though the dust optical depth is small, the high luminosity of the star ($\textgreater$ 10$^5$ $L_\mathrm{\odot}$) results in a significant surface brightness of the dust at mid-IR wavelengths ($\sim$ 10 MJy/sr) which is easily detectable with IR instrumentation such as IRAS, Spitzer and more recently, WISE.

$\sigma$ Ori AB is a suitable candidate to observe a dust wave. First, the ionized flow from the L1630 molecular cloud creates a smooth homogeneous background without interfering factors which are often seen in regions where radiation pressure is important. For example, mass-loss variability in AGB stars or more complex geometries of compact \HII regions will confuse a dust wave or bow wave with its surroundings. In the special case of $\sigma$ Ori AB, these disturbing factors are minimal and the dust wave can be seen in great contrast against the underlying flow. Second, $\sigma$ Ori AB has been classified as a weak-wind candidate \citep{najarro_2011}. According to this scenario, the bow shock stand-off distance should lie at $r_\mathrm{s}$ = 8 $\times$ 10$^{-3}$ pc ($\sim$ 4$\arcsec$) (see Sec. \ref{sec:bowshock}). If $\sigma$ Ori AB would have a powerful stellar wind according to wind parameters derived by \citet{howarth_1989}, the bow shock stand-off distance would lie at a greater distance from the star ($r_\mathrm{s}$ = 0.13 pc). In this case, incoming dust from the flow might be shocked by the stellar wind before it would reach the dust wave zone. 

We argue that dust waves and bow waves should be common around stars with weak winds. Accurate wind parameters are still scarce for low mass loss rates, but it has become clear that stars with log($L/L_\odot$) $\textless$ 5.2 show the weak-wind phenomenon. Although dust waves and bow waves are more likely to be formed around lower mass stars moving slowly through the ISM, only the most massive stars will create a structure which has a large enough separation from the star to resolve at IR wavelengths. Valuable information on dust properties can be probed by studying the size and geometry of a dust waves and bow waves. This could give a handle in studying the properties of dust inside different phases of the ISM and \HII regions, which remain until now poorly understood.

\begin{acknowledgements} 
Studies of interstellar dust and chemistry at Leiden Observatory are supported through advanced ERC grant 246976 from the European Research Council, through a grant by the Dutch Science Agency, NWO, as part of the Dutch Astrochemistry Network, and through the Spinoza premie from the Dutch Science Agency, NWO. NLJC acknowledges support from the Belgian Federal Science Policy Office via the PRODEX Programme of ESA.
 \end{acknowledgements}

\bibliographystyle{aa} 
\bibliography{sigmaori.bib} 

\appendix
\section{IR arcs around candidate weak-wind stars}\label{sec:appendix}

In Tab. \ref{tab:weakir}, we list parameters necessary for the calculation of the location of a bow shock ($r_\m{s}$) or dust wave/bow wave ($r_\m{min}$) for the weak-wind candidates around which an IR-arc is observed. We calculate the stand-off distance both for the weak-wind scenario ($r_\m{s,weak}$) and the normal wind scenario ($r_\m{s,Vink}$). In most cases, the local ISM density is taken from references in the literature. However, for HD48279 and HD216898 these values are not directly measured. For these sources we used the average density along the line of sight, which may not be representable for the local density. $r_\m{min}$ is calculated with the model presented in this paper for a streamline with impact parameter $b$ = 0. The grains approach the star with velocity $v_\m{\star}$ from infinity (i.e., a large radius compared to the distance travelled during the lifetime of the star) and are stopped in front of the star at $r_\m{min}$. The listed $r_\m{min}$ contains two different values, where the lower value corresponds to 3000 \AA\ silicate grains and the higher value to 300 \AA\ silicate grains. Grains with sizes between these values are stratified in a region bounded by the listed values.

Using weak wind parameters, the calculated stand-off distance $r_\m{s,weak}$ is well below $r_\m{obs}$. The other goes for the stand-off distance $r_\m{s,Vink}$ using the theoretical recipe from \citet{vink_2000}: except for $\sigma$ Ori AB, $r_\m{s,Vink}$ exceeds $r_\m{obs}$. We conclude that in order to reproduce the IR arcs at $r_\m{obs}$, the stars need to have stellar winds close to the normal wind scenario. In the dust wave scenario, the region bounded by the 3000 \AA\ and 300 \AA\ grains encapsulates the observed peak distance of the arc structure $r_\m{obs}$, and can explain the IR structures at the observed locations. HD34078 is most likely a special case; neither a stellar wind nor radiation pressure can explain the observed location of the peak IR emission, due to the extreme high velocity of the star and the high density of the molecular cloud or core which it recently encountered. More observations are required to further constrain the relevant physical parameters of this system.   

\newpage

\begin{sidewaystable*}
\vspace{18cm}
\centering
\begin{tabular}{|l|l|l|l|l|l|l|l|l|l|l|l|l|l|l|}\hline
Star & Sp. T & $d$ & $N_\m{H}$ & $n_\m{H}$ & $n$ & $v_{\star}$ & $v_\infty$ & $\dot{M}_\m{weak}$ & $\dot{M}_\m{Vink}$  & $r_\m{obs}$ & $r_\m{s,weak}$ & $r_\m{s,Vink}$ & $r_\m{min}$ \\ 
& & (kpc) & (cm$^{-2}$) &  (cm$^{-3}$) &  (cm$^{-3}$) & (km s$^{-1}$)  & (km s$^{-1}$)  & (M$_\odot$ yr$^{-1}$)  & (M$_\odot$ yr$^{-1}$) & (pc) & (pc) & (pc) & (pc) \\ \hline
HD34078 & O9.5V$^{(1)}$ & 0.4 & 1.8e21 & 1.4 & 2e3$^{(3), \star}$ & 150$^{(5)}$ & 800$^{(1)}$ & 3.2e-10$^{(1)}$ & 4.2e-8$^{(1)}$ & 0.06 & 1.7e-4 & 1.9e-3 & 5e-3-8e-3 \\
HD48099 & O5.5$^{(2)}$ & 1.5 & 1.4e21 & 0.35 & ~1$^{(4)}$ & 39$^{(5)}$ & 2800$^{(2)}$ & 2.5e-8$^{(2)}$ & 2.1e-6$^{(2)}$ & 1.85 & 0.48 & 4.49 & 0.52-2.73 \\ 
HD48279 & O8.5V$^{(2)}$ & 1.5 & 2.0e21 & 0.45 & ... & 30$^{(6)}$ & 1700$^{(2)}$ & 1.6e-9$^{(2)}$ & 1.6e-7$^{(2)}$ & 0.43  & 0.19 & 1.73 & 0.22-1.73 \\
HD149757 & O9.5V & 0.1 & 5.0e20 & 1.15 & 4$^{(7)}$ & 26.5 & 1500 & 1.6e-9$^{(7)}$ & 1.3e-7$^{(7)}$ & 0.16 & 0.07 & 0.60 & 0.17-0.80  \\
HD216898 & O9V$^{(8)}$ & 0.6 & 1.1e21$^{\star \star}$ & 0.60$^{\star \star}$ & ... & 24.8 & 1700$^{(8)}$ & 4.5e-10$^{(8)}$ & 6.0e-8$^{(8)}$ & 0.56 & 0.10 & 1.98 & ... \\ 
$\sigma$ Ori AB & O9.5V & 0.3 & ... & ... & 10 & 50 & 1500 & 2.0e-10$^{(9)}$ & 8.0e-8$^{\star \star \star}$ & 0.1 & 6e-3 & 0.13 & 0.10-0.40 \\ \hline
\end{tabular}
\caption{IR arcs around candidate weak-wind stars. Column definitions: $d$ is the distance; $N_\m{H}$ and $n_\m{H}$ are the total column density and average number density of atomic hydrogen along of sight; $n$ is a directly measured local ISM number density (see below); $\dot{M}_\m{weak}$ and $\dot{M}_\m{Vink}$ are the weak-wind values as given in the corresponding references and the mass-loss as predicted by the theoretical recipe described in \citet{vink_2000}; $r_\m{obs}$ is the observed projected distance where the emission peaks in the WISE 22 $\mu$m images (assuming the listed $d$); $r_\m{s,weak}$, $r_\m{s,Vink}$ are the calculated stand-off distance using $\dot{M}_\m{weak}$, $\dot{M}_\m{Vink}$ and $v_\infty$; $r_\m{min}$ is the predicted location of the dust wave in the model described in this work. $r_\m{min}$ calculated through spectral type, density, velocity, and listed for 3000 \AA\ and 300 \AA\ grains, respectively. References: (1) \citet{martins_2005}; (2) \citet{martins_2012}; (3) \citet{boisse_2009}; (4) \citet{brown_2005}; (5) \citet{peri_2012}; (6) \citet{gvaramadze_2012b}; (7) \citet{gvaramadze_2012}; (8) \citet{marcolino_2009}; (9) \citet{najarro_2011}. Distances and (average) column densities from \citet{gudennavar_2012} (for HD216898 $d$ was not listed and is calculated through the measured {\em Hipparcos} parallax from \citet{van_leeuwen_2007}). $^{\star}$: HD34078 is a fast runaway star which has recently encountered a molecular cloud or core not along the line of sight, therefore the average density along the line of sight is not representable in calculating the stand-off distance. $^{\star \star}$: There is no column density of atomic hydrogen measured for HD216898; the listed column density (and the derived average density) is for molecular hydrogen. $^{\star \star \star}$: For consistency, we list $\dot{M}$ from \citet{howarth_1989}, value used throughout this work. This value is consistent with the \citet{vink_2000} prediction, seen by comparison with the other O9.5V stars HD34078 and HD149757.}
\label{tab:weakir}
\end{sidewaystable*}

\end{document}